\documentclass{emulateapj}

\usepackage{apjfonts}
\newcommand{\etal}{et al.}
\newcommand{\CIV}{C{\sevenrm IV}}
\newcommand{\SiIV}{Si{\sevenrm IV}}
\newcommand{\CIII}{C{\sevenrm III]}}
\newcommand{\MgII}{Mg{\sevenrm II}}
\newcommand{\OIV}{O{\sevenrm IV}}
\newcommand{\bracket}[1]{\left\langle#1\right\rangle}
 
 \font\sevenrm=cmr7 scaled 1000

\slugcomment{Submitted to AJ on 3 November 2006}

\begin{document}
\title{Clustering of High Redshift ($z\ge 2.9$) Quasars from the
Sloan Digital Sky Survey}

\shorttitle{Quasar Correlation Function at $z\ge 2.9$}

\shortauthors{Y. Shen et al.}

\author{Yue Shen\altaffilmark{1}, Michael A. Strauss\altaffilmark{1},
  Masamune Oguri\altaffilmark{1,2}, Joseph F. Hennawi\altaffilmark{3},
  Xiaohui Fan\altaffilmark{4}, Gordon T. Richards\altaffilmark{5},
  Patrick B. Hall\altaffilmark{6}, James E. Gunn\altaffilmark{1}, Donald
  P. Schneider\altaffilmark{7}, Alexander S. Szalay\altaffilmark{8},
  Anirudda R. Thakar\altaffilmark{8}, Daniel E. Vanden
  Berk\altaffilmark{7}, Scott F. Anderson\altaffilmark{9}, Neta
  A. Bahcall\altaffilmark{1}, Andrew J. Connolly\altaffilmark{10}, Gillian R. Knapp\altaffilmark{1}}

\altaffiltext{1}{Princeton University Observatory, Princeton, NJ 08544.}

\altaffiltext{2}{Kavli Institute for Particle Astrophysics and Cosmology, Stanford
   University, 2575 Sand Hill Road, Menlo Park, CA 94025. }

\altaffiltext{3}{Department of Astronomy, Campbell Hall, University of California,
  Berkeley, California 94720}

\altaffiltext{4}{Steward Observatory, 933 North Cherry Avenue, Tucson, AZ 85721.}

\altaffiltext{5}{Department of Physics, Drexel University, 3141 Chestnut Street, Philadelphia, PA 19104.}

\altaffiltext{6}{Dept. of Physics \& Astronomy, York University,
4700 Keele St., Toronto, ON, M3J 1P3, Canada.}

\altaffiltext{7}{Department of Astronomy and Astrophysics, 525 Davey Laboratory, Pennsylvania State
University, University Park, PA 16802.}

\altaffiltext{8}{Center for Astrophysical Sciences, Department of Physics and Astronomy, Johns
Hopkins University, 3400 North Charles Street, Baltimore, MD 21218.}

\altaffiltext{9}{Department of Astronomy, University of Washington, Box 351580, Seattle, WA
98195.}

\altaffiltext{10}{Department of Physics and Astronomy, University
of Pittsburgh, 3941 O'Hara Street, Pittsburgh, PA 15260.}

\begin{abstract}
We study the two-point correlation function of a uniformly
selected sample of 4,426 luminous optical quasars with redshift
$2.9 \le z\le 5.4$ selected over 4041 deg$^2$ from the Fifth Data
Release of the Sloan Digital Sky Survey.  We fit a power-law to
the projected correlation function $w_p(r_p)$ to marginalize over
redshift space distortions and redshift errors. For a real-space
correlation function of the form $\xi(r)=(r/r_0)^{-\gamma}$, the
fitted parameters in comoving coordinates are $r_0 = 15.2 \pm
2.7\, h^{-1}$ Mpc and $\gamma = 2.0 \pm 0.3$, over a scale range
$4\le r_p\le 150\ h^{-1}$ Mpc. Thus high-redshift quasars are
appreciably more strongly clustered than their $z \approx 1.5$
counterparts, which have a comoving clustering length $r_0 \approx
6.5\ h^{-1}$ Mpc. Dividing our sample into two redshift bins:
$2.9\le z\le 3.5$ and $z\ge 3.5$, and assuming a power-law index
$\gamma=2.0$, we find a correlation length of $r_0 = 16.9 \pm
1.7\,h^{-1}$ Mpc for the former, and $r_0 = 24.3 \pm 2.4\,h^{-1}$
Mpc for the latter. Strong clustering at high redshift indicates
that quasars are found in very massive, and therefore highly
biased, halos. Following Martini \& Weinberg, we relate the
clustering strength and quasar number density to the quasar
lifetimes and duty cycle. Using the Sheth \& Tormen halo mass
function, the quasar lifetime is estimated to lie in the range
$4\sim 50$ Myr for quasars with $2.9\le z\le 3.5$; and $30\sim
600$ Myr for quasars with $z\ge 3.5$. The corresponding duty
cycles are $0.004\sim 0.05$ for the lower redshift bin and
$0.03\sim 0.6$ for the higher redshift bin. The minimum mass of
halos in which these quasars reside is $2-3\times 10^{12}\
h^{-1}M_\odot$ for quasars with $2.9\le z\le 3.5$ and $4-6\times
10^{12}\ h^{-1}M_\odot$ for quasars with $z\ge 3.5$; the effective
bias factor $b_{\rm eff}$ increases with redshift, e.g., $b_{\rm
eff}\sim 8$ at $z=3.0$ and $b_{\rm eff}\sim 16$ at $z=4.5$.
\end{abstract}

\section{Introduction}
Recent galaxy surveys (e.g., the 2dF Galaxy Redshift Survey,
Colless \etal\ 2001 and the Sloan Digital Sky Survey (SDSS), York
\etal\ 2000) have provided ample data for the study of the
large-scale distribution of galaxies in the present-day Universe.
The clustering of galaxies, which are tracers of the underlying
dark matter distribution, gives a powerful test of hierarchical
structure formation theory, especially when compared with
fluctuations in the Cosmic Microwave Background.  Indeed, the
results show excellent agreement with the now-standard flat
$\Lambda$-dominated concordance cosmology (e.g., Spergel \etal\
2003, 2006; Tegmark \etal\ 2004, 2006; Eisenstein \etal\ 2005;
Percival \etal\ 2006). The galaxy two-point correlation function
is well-fit by a power law:
$\xi(r)=(r/r_0)^{-\gamma}$ on scales $r\lesssim 20$ $h^{-1}$ Mpc,
with comoving correlation length $r_0\sim 5$ $h^{-1}$ Mpc and
slope $\gamma\sim 1.8$ (Totsuji \& Kihara 1969; Groth \& Peebles
1977; Davis \& Peebles 1983; Hawkins \etal\ 2003), although there
is an excess above the power law below $2\,h^{-1}$ Mpc, thought to
be due to halo occupation effects (Zehavi \etal\ 2004, 2005).

At high redshifts and earlier times, the dark matter clustering
strength should be weaker, but the first clustering studies of
high-redshift galaxies with the Keck telescope (Cohen \etal\ 1996;
Steidel \etal\ 1998; Giavalisco \etal\ 1998; Adelberger \etal\
1998) showed that galaxies at $z > 3$ show a similar comoving
correlation length to those of today, results that have since been
confirmed with much larger samples (e.g., Adelberger \etal\ 2005a;
Ouchi \etal\ 2005; Kashikawa \etal\ 2006; Meneux \etal\ 2006; Lee
\etal\ 2006; Quadri \etal\ 2006). This is indeed expected:
high-redshift galaxies are thought to form at rare peaks in the
density field, which will be strongly biased relative to the dark
matter (Kaiser 1984; Bardeen \etal\ 1986); under gravitational
instability, the bias of galaxies drops over time as a function of
redshift (Tegmark \& Peebles 1998; Blanton \etal\ 2000; Weinberg
\etal\ 2004).

Luminous quasars offer a different probe of the clustering of
galaxies at high redshift.  Powered by gas accretion onto central
super-massive black holes (Salpeter 1964; Lynden-Bell 1969),
quasars are believed to be the progenitors of local dormant
super-massive black holes which are ubiquitous in the centers of
nearby bulge-dominated galaxies (e.g., Kormendy \& Richstone 1995;
Magorrian \etal\ 1998; Yu \& Tremaine 2002).  Studies of the
clustering properties of quasars date back to Osmer (1981); in
general, quasars have a clustering strength similar to that of
luminous galaxies at the same redshift (Shaver 1984; Croom \&
Shanks 1996; Porciani, Magliocchetti \& Norberg 2004, hereafter
PMN04; Croom \etal\  2005).  If the triggering of quasar activity
is not tied to the larger-scale environment in which their host
galaxies reside, this is not a surprising result; quasars are
interpreted as a stochastic process through which every luminous
galaxy passes, and therefore the clustering of quasars should be
no different from that of luminous galaxies. Studies of the
clustering of galaxies around quasars similarly find that quasar
environments are similar to those of luminous galaxies (Serber \etal\
2006, and references therein), although evidence for an
enhanced clustering of quasars on small scales (Djorgovski 1991;
Hennawi \etal\ 2006a; but see also Myers \etal\ 2006c) suggests
that tidal effects within 100 kpc may trigger quasar activity.

A number of studies have examined the redshift evolution of quasar
clustering, but the results have been controversial: some papers
conclude that quasar clustering either decreases or weakly evolves
with redshift (e.g., Iovino \& Shaver 1988; Croom \& Shanks 1996),
while others say that it increases with redshift (e.g., Kundic
1997; La Franca \etal\ 1998; PMN04; Croom \etal\ 2005).  Myers
\etal\ (2006a, b, c) examined the clustering of quasar candidates
with photometric redshifts from the SDSS; they find little
evidence for evolution in clustering strength between $z \approx
2$ and today.
These studies also find little evidence for a strong luminosity
dependence of the quasar correlation function (e.g., Croom \etal\
2005; Connolly \etal, in preparation), which is in accord with
quasar models in which quasar luminosity is only weakly related to
black hole mass (Lidz \etal\ 2006).

The vast majority of quasars in flux-limited samples like the SDSS
(and especially UV-excess surveys like the 2dF QSO Redshift
Survey; Croom \etal\ 2004) are at relatively low redshift, $z <
2.5$.  More distant quasars are intrinsically rarer (e.g.,
Richards \etal\ 2006), and at a given luminosity are of course
substantially fainter. However, we might expect high-redshift
quasars to be appreciably more biased than their lower-redshift
counterparts.  The high-redshift quasars in flux-limited samples
are intrinsically luminous, and by the Eddington argument, are
powered by massive ($>10^8\,M_\odot$) black holes.  If the
relation between black hole mass and bulge mass (Tremaine \etal\
2002 and references therein), and by extension, black hole mass
and dark matter halo mass (Ferrarese 2002) holds true at high
redshift, then luminous quasars reside in very massive, and
therefore very rare halos at high redshift.  Rare, many$-\sigma$
peaks in the density field are strongly biased (Bardeen \etal\
1986).  Thus detection of particularly strong clustering at high
redshift would allow tests both of the relationship between
quasars and their host halos, and the predictions of biasing
models. The rarity of the halos in which quasars reside is of
course related to the observed number density of quasars and their
duty cycle/lifetime, thus the quasar luminosity function and the
quasar clustering properties can be used to constrain the average
quasar lifetime $t_{\rm Q}$ (Haiman \& Hui 2001; Martini \&
Weinberg 2001), or equivalently, the duty cycle: the fraction of
time a supermassive black hole shines as a luminous quasar.

Studies to date of the clustering of high-redshift quasars have
been hampered by small number statistics.  Stephens \etal\ (1997)
and Kundic (1997) examined three $z> 2.7$ quasar pairs with
comoving separations $5-10$ $h^{-1}$ Mpc in the Palomar Transit
Grism Survey of Schneider \etal\ (1994), and estimated a comoving
correlation length $r_0\sim 17.5\pm 7.5$ $h^{-1}$ Mpc, which is
three times higher than that of lower redshift quasars.  Schneider
\etal\ (2000) found a pair of $z = 4.25$ quasars in the SDSS
separated by less than $2\ h^{-1}$ Mpc; this single pair implies a
lower limit to the correlation length of $r_0 = 12\ h^{-1}$ Mpc.
Similarly, the quasar pair separated by a few Mpc at $z \sim 5$
found by Djorgovski \etal\ (2003) also implies strong clustering
at high redshift. However, measuring a true correlation function
requires large samples of quasars.  At $z \sim 4$, the mean
comoving distance between luminous ($M_i < -27.6$) quasars is
$\sim 150\ h^{-1}$ Mpc (Fan \etal\ 2001; Richards \etal\ 2006),
thus to build up statistics on smaller-scale clustering in such a
sparse sample requires a very large volume.  The SDSS quasar
sample is the first survey of high-redshift quasars that covers
enough volume to allow this measurement to be made.

This paper presents the correlation function of high redshift
($z\ge 2.9$) quasars using the fifth data release (DR5;
Adelman-McCarthy \etal\ 2007) of the SDSS.  DR5 contains $\sim
6,000$ quasars with redshift $z\ge 2.9$.  We construct a
homogeneous flux-limited sample for clustering analysis in
\S~\ref{sec:sample}, with special focus on redshift determination
in Appendix~\ref{app:redshift}, and the angular mask of the sample
in Appendix~\ref{app:geometry}. We present the correlation
function itself in \S~\ref{sec:cf}, together with a discussion of
its implications for quasar duty cycles and lifetimes.  We
conclude in Section \ref{sec:con}. Throughout the paper we use the
third year WMAP $+$ all parameters\footnote{{\it
http://lambda.gsfc.nasa.gov/product/map/current/params/lcdm\_all.cfm}}
(Spergel \etal\ 2006) for the cosmological model: $\Omega_M=0.26$,
$\Omega_{\Lambda}=0.74$, $\Omega_b=0.0435$, $h=0.71$, $n_s=0.938$,
$\sigma_8=0.751$. Comoving units are used in distance
measurements; for comparison with previous results, we will often
quote distances in units of $h^{-1}$ Mpc.

\section{Sample Selection}
\label{sec:sample}
\subsection{The SDSS Quasar Sample}\label{sec:parent_sample}
The SDSS uses a dedicated 2.5-m wide-field telescope (Gunn \etal\
2006) which uses a drift-scan camera with 30 $2048 \times 2048$
CCDs (Gunn \etal\ 1998) to image the sky in five broad bands
($u\,g\,r\,i\,z$; Fukugita \etal\ 1996).  The imaging data are
taken on dark photometric nights of good seeing (Hogg \etal\
2001), are calibrated photometrically (Smith \etal\ 2002; Ivezi\'c
\etal\ 2004; Tucker \etal\ 2006) and astrometrically (Pier \etal\
2003), and object parameters are measured (Lupton \etal\ 2001;
Stoughton \etal\ 2002). Quasar candidates (Richards \etal\ 2002b)
for follow-up spectroscopy are selected from the imaging data
using their colors, and are arranged in spectroscopic plates
(Blanton \etal\ 2003) to be observed with a pair of double
spectrographs. The quasars observed through the Third Data Release
(Abazajian \etal\ 2005) have been cataloged by Schneider \etal\
(2005), while Schneider \etal\ (2006) extend this catalog to the
DR5.  In this paper, we will use results from DR5, for which
spectroscopy has been carried out over 5740 deg$^2$. Because of
the diameter of the fiber cladding, two targets on the same plate
cannot be placed closer than $55''$ (corresponding to $\sim 1.2\
h^{-1}$ Mpc at $z=3$)\footnote{Serendipitous objects closer than
$55''$ might be observed on overlapped plates.}; the present paper
therefore concentrates on clustering on larger scales, and we will
present a discussion of the correlation function on small scales
in a paper in preparation.

The quasar target selection algorithm is in two parts: quasars
with $z \le 3.5$ are outliers from the stellar locus in the $ugri$
color cube, while those with $z > 3.5$ are selected as outliers in
the $griz$ color cube.  The quasar candidate sample is
flux-limited to $i = 19.1$ (after correction for Galactic
extinction following Schlegel, Finkbeiner, \& Davis 1998), but
because high-redshift quasars are quite rare, the magnitude limit
for objects lying in those regions of color space corresponding to
quasars at $z > 3$ are targeted to $i = 20.2$.  The quasar locus
crosses the stellar locus in color space at $z \approx 2.7$ (Fan
1999), meaning that quasar target selection is quite incomplete
there (Richards \etal\ 2006). For this reason, we have chosen to
define high-redshift quasars as those with $z \ge 2.9$.

We draw our parent sample from the SDSS DR5 catalog.  We have
taken all quasars with listed redshift $z \ge 2.9$ from the DR3
quasar catalog (Schneider \etal\ 2005); the redshifts of these
objects have all been checked by eye, and we rectify a small
number of incorrect redshifts in the database.  This sample
contains 3,333 quasars. In addition, we have included all objects
on plates taken since DR3 with listed redshift $z \ge 2.9$ as
determined either from the official spectroscopic pipeline which
determines redshifts by measuring the position of emission lines
(SubbaRao \etal\ 2002) or an independent pipeline which fits
spectra to quasar templates (Schlegel \etal, in preparation).  We
examined by eye the spectra of all objects with discrepant
redshifts between the two pipelines. There are 2,805 quasars added
to our sample from plates taken since DR3.

Quasar emission lines are broad, and tend to show systematic
wavelength offsets from the true redshift of the object (Richards
\etal\ 2002a and references therein).  Appendix~\ref{app:redshift}
describes our investigation of these effects, determination of an
unbiased redshift for each object, and the definition of our final
sample of 6,109 quasars with $z \ge 2.9$ (after rejecting 29
objects that turn out to have $z<2.9$).

\subsection{Clustering Subsample}
\label{sec:cluster_sample}

Not all the quasars in our sample are suitable for a clustering
analysis.  Here we follow Richards \etal\ (2006) and select only
those quasars that are selected from a uniform algorithm.  In
particular:
\begin{itemize}
\item The version of the quasar target selection algorithm used
for the SDSS Early Data Release (Stoughton \etal\ 2002) and the
First Data Release (DR1; Abazajian \etal\ 2003) did a poor job of
selecting objects with $z \approx 3.5$.  We use only those quasars
targeted with the improved version of the algorithm, i.e., those
with target selection version no lower than v3\_1\_0.

\item Some quasars are found using algorithms other than the
quasar target selection algorithm described by Richards \etal\
(2002b), including special selection in the Southern Galactic Cap
(see Adelman-McCarthy \etal\ 2006) and optical counterparts to
ROSAT sources (Anderson \etal\ 2003).  The completeness of these
auxiliary algorithms is poor, and we only include quasars targeted
by the main algorithm.

\item Because quasars are selected by their optical colors,
regions of sky in which the SDSS photometry is poor are unlikely
to have complete quasar targeting.
\end{itemize}

We now describe how the regions with poor photometry are
identified.
The SDSS images are processed in a series of $10'\times 13'$ {\it
fields}.  We follow Richards \etal\ (2006) and mark a given field has
having bad photometry if any one of the following criteria is
satisfied:
\begin{itemize}
\item the $r$-band seeing is greater than 2$''$.0; \item The
operational database quality flag for that field is BAD, MISSING
or HOLE (only 0.15\% of all DR5 fields have one of these flags set);
\item
The median difference between the PSF and large-aperture
photometry magnitudes of bright stars lies more than 3$\sigma$
from the mean over the entire DR5 sample in any of the five bands;

\item Any of the four principal colors of the stellar locus
(Ivezi\'c \etal\ 2004) deviates from the mean of the DR5 sample by
more than 3$\sigma$;

\item Any of the four values of the rms scatter around the mean
principal color deviates from the mean over DR5 by more than
5$\sigma$, or, deviates from the DR5 mean by more than 2$\sigma$,
and also deviates from the mean of that {\em run} by more than
3$\sigma$. This criterion reflects the fact that the statistics of
the rms widths of the principal color distributions per field vary
significantly from run to run.
\end{itemize}
All the information we need to identify bad fields in this way can
be retrieved from the {\tt runQA} table in the SDSS Catalogue
Archive Server (CAS\footnote{\tt http://cas.sdss.org}). A total of
13.24\% of the net area of the clustering subsample is marked as
bad. These bad fields will serve as a secondary mask in our
geometry description. We will compute the correlation function
both including and excluding the bad regions, to test our
sensitivity to possible selection problems in the bad regions.

Finally, due to overlapping plates, there are roughly 200 duplicate
objects in our parent sample, which we identified and removed using
objects' positions.

Our final cleaned subsample contains 4,426 quasars before
excluding bad fields and 3,846 quasars with bad fields excluded.
Thus 13.1\% of high-redshift quasars are in bad fields,
essentially identical to the fraction of the area flagged as bad,
which suggests that the selection of quasars in these regions is
not terribly biased. A list of the unique high-redshift quasars in
our parent sample and in the subsample used in our clustering
analysis is provided in Table~\ref{table:highz_sub}.

\begin{deluxetable*}{lcccrccccc}
\tablecolumns{10} \tablewidth{\textwidth} \tablecaption{High
redshift quasar sample\label{table:highz_sub}} \tablehead{Plate &
Fiber & MJD & RA (deg) & DEC (deg)  & $z$ & $z_{\rm err}$ & $i$
mag & sub\_flag & good\_flag} \startdata
1091 & 553 & 52902  &  0.193413 & 1.239112 & 3.741 & 0.011   &19.74 & 0 & 0\\
1489 & 506 & 52991  &  0.214856 & 0.200710 & 3.881 & 0.030   &19.97 & 0 & 0\\
1489 & 104 & 52991  &  0.397978 &$-0.701886$ & 3.572 & 0.008 &19.33 & 0 & 0\\
0387 & 556 & 51791  &  0.587972 & 0.363741 & 3.057 & 0.010   &18.58 & 0 & 0\\
0650 & 111 & 52143  &  0.660070 &$-10.197168$& 3.942 & 0.012 &19.97 & 0 & 0\\
0750 & 608 & 52235  &  0.751425 & 16.007709& 3.689 & 0.011   &19.50 & 1 & 1\\
0650 & 048 & 52143  &  0.763943 &$-10.864079$& 3.645 & 0.011 &19.20 & 0 & 0\\
0750 & 036 & 52235  &  0.896718 & 14.795454& 3.462 & 0.012   &19.95 & 1 & 1\\
0750 & 632 & 52235  &  1.155146 & 15.174562& 3.203 & 0.009   &20.17 & 1 & 1\\
0751 & 207 & 52251  &  1.401625 & 13.997071& 3.705 & 0.011   &19.34 & 1 & 1\\
\enddata
\tablecomments{\footnotesize The entire high redshift quasar
sample with duplicate objects removed. The {\em sub\_flag} is 1
when an object is in the clustering subsample, and the {\em
good\_flag} is 1 for objects lying in good fields.  The $i$
magnitudes are SDSS PSF (asinh) magnitudes corrected for Galactic
extinction (Schlegel \etal\ 1998); they use the ubercalibration
described by Padmanabhan \etal\ (2007), which differs slightly
from that used in the official DR5 quasar catalog (Schneider
\etal\ 2007). The entire table is available in the electronic
edition of the paper.}
\end{deluxetable*}

\subsection{Distribution of Quasars in Angle and on the Sky}
\label{sec:distribution}

\begin{figure*}
  \centering
    \includegraphics[width=\textwidth, height=0.5\textwidth]{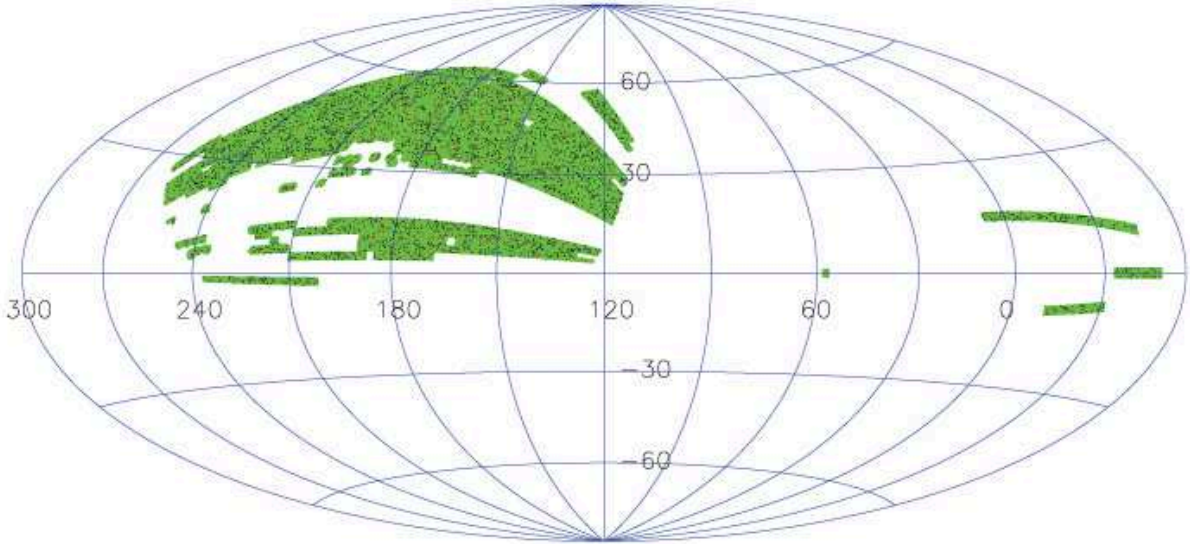}
    \caption{Aitoff projection in equatorial coordinates of the
      angular coverage of our clustering subsample (with all fields).
    The center of the plot is the direction $\hbox{RA}=120^\circ$ and $\hbox{Dec}=0^\circ$. The
    dots indicate quasars in our clustering subsample, with red dots indicating those in bad imaging fields.
    The angular coverage is patchy due to the various selection criteria described in \S\ref{sec:cluster_sample} and
    Appendix \ref{app:geometry}.  For example, much of the Southern
    Equatorial Stripe ($\delta = 0$, $300 < \alpha < 60^\circ$) was targeted using the old version of the quasar
    targetting algorithm.}
    \label{fig:sky_cover}
\end{figure*}

\begin{figure}
  \centering
    \includegraphics[width=0.50\textwidth]{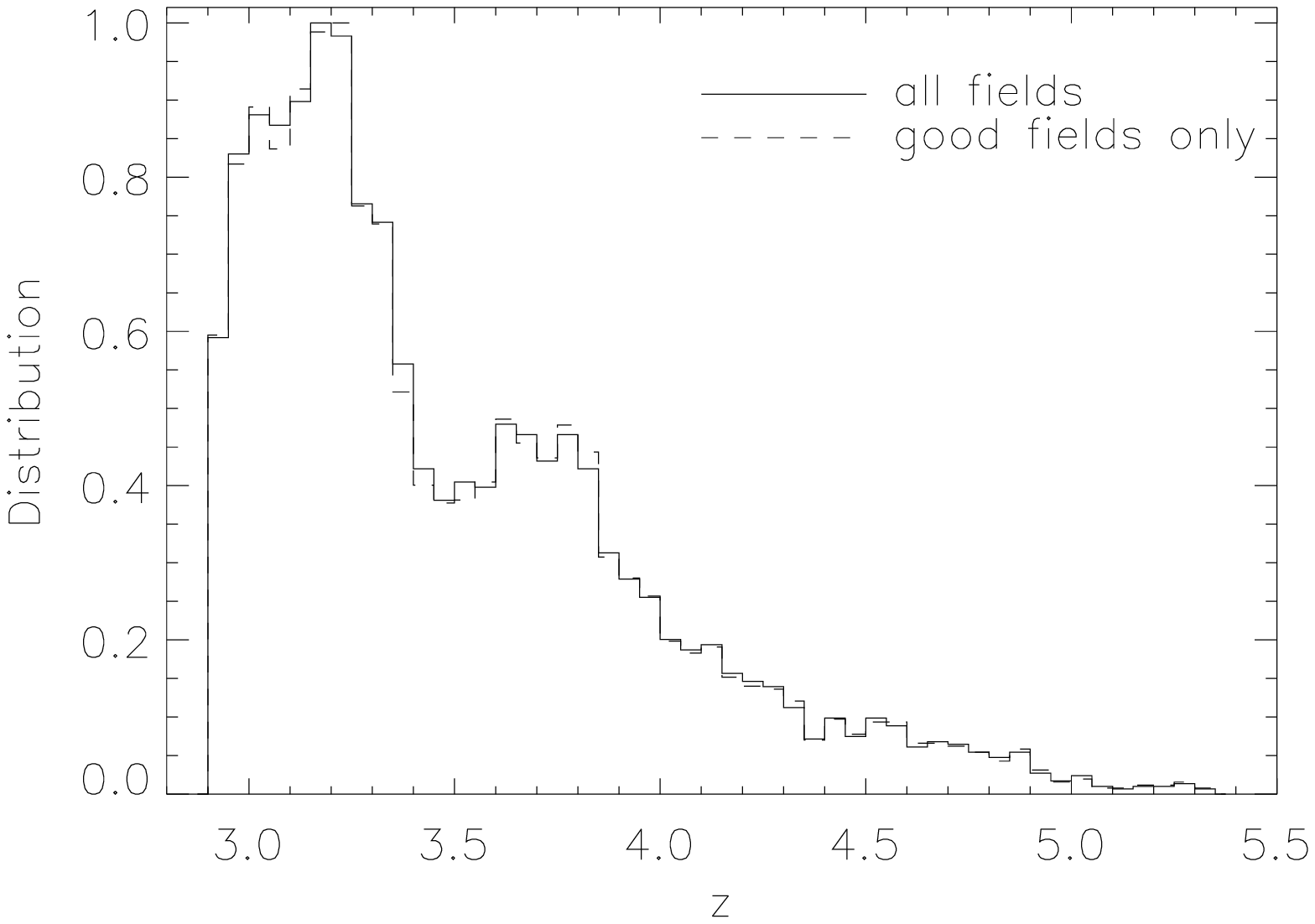}
    \caption{Observed redshift distribution of our quasar clustering subsamples, normalized by
    the peak value.  This distribution is the product of the evolution of the quasar
    density distribution, and the quasar selection function; the
    latter is responsible for the dip at $z \thickapprox 3.5$, where
    quasars have very similar colors to those of G and K stars.  We
    show the redshift distributions for the subsamples both including
    and excluding bad fields; the results are essentially identical.
    The redshift binning is $\Delta z = 0.05$.
}
    \label{fig:z_distri}
\end{figure}

The footprint of our quasar clustering subsample is quite
complicated. The definition of the sample's exact boundaries,
needed for the correlation function analysis which follows, is
described in detail in Appendix~\ref{app:geometry}.
Fig.~\ref{fig:sky_cover} shows the area of sky from which the
sample was selected in green, and the sample of quasars is
indicated as dots, with red dots indicating objects in bad imaging
fields. The total area subtended by the sample is 4041 deg$^2$;
when bad fields are excluded, the solid angle drops to 3506
deg$^2$.

The target selection algorithm for quasars is not perfect and the
selection function depends on redshift.  Our sample is limited to
$z \ge 2.9$; at slightly lower redshift, the broad-band colors of
quasars are essentially identical to those of F stars (Fan 1999),
giving a dramatic drop in the quasar selection function. Moreover,
as discussed in Richards \etal\ (2006), quasars with redshift
$z\thickapprox 3.5$ have similar colors to G/K stars in the $griz$
diagram and hence targeting becomes less efficient around this
redshift (as mentioned above, this problem was even worse for the
version of target selection used in the EDR and DR1).  This is
reflected in the redshift distribution of our sample
(Fig.~\ref{fig:z_distri}), which shows a dip at $z \thickapprox
3.5$.  We will use these distributions in computing the
correlation function below.

\section{Correlation Function}\label{sec:cf}
Now that we understand the angular and radial selection function
of our sample, we are ready to compute the two-point correlation
function.  Doing so requires producing a random catalog of points
(i.e., without any clustering signal) with the same spatial
selection function.  We will first compute the correlation
function in ``redshift space''
in \S~\ref{sec:redshift_space}, then
derive the real-space correlation function in
\S~\ref{sec:real_space} by projecting over redshift space
distortions.  Our calculations will be done both including and
excluding the bad fields (\S~\ref{sec:cluster_sample}); we will
find that our results are robust to this detail.

\subsection{``Redshift Space'' Correlation Function}
\label{sec:redshift_space}

\begin{deluxetable}{lrrrrr}
\tablecolumns{6}\tablewidth{0pt} \tablecaption{Redshift space
correlation function $\xi_s(s)$\label{table:xi_s}} \tablehead{$s$
($h^{-1}$ Mpc)
 & ${\rm DD_{mean}}$  & ${\rm RR_{mean}}$&${\rm
DR_{mean}}$&$\xi_s$ & $\xi_s$ error } \startdata
2.244 & 0.0 & 0.9 & 0.0 & -- & -- \\
2.825 & 0.0 & 5.4 & 0.0 & -- & -- \\
3.557 & 0.0 & 6.3 & 0.0 & -- & -- \\
4.477 & 1.8 & 14.4 & 0.9 & 16.5 & 12.8 \\
5.637 & 0.0 & 34.2 & 3.6 & -- & -- \\
7.096 & 1.8 & 38.7 & 11.7 & 3.54 & 3.61 \\
8.934 & 1.8 & 99.0 & 18.0& 1.26 & 1.88 \\
11.25 & 2.7 &215.0 &36.9 & 0.663 & 0.733 \\
14.16 & 4.5 &406.5&80.0 & 0.191 &0.786 \\
17.83 & 8.9 &804.2&162.4 & 0.131 &0.472 \\
22.44 &15.2 &1592.4&279.4& 0.236 &0.175 \\
28.25 &22.4 &3123.6 & 607.3& $-$0.280& 0.223\\
35.57 &70.7 &6028.6 & 1139.3& 0.361& 0.170\\
44.77 &104.9 &11959.1 & 2137.1& 0.101& 0.121\\
56.37 &210.9&23480.2& 4381.2& 0.0384& 0.0862\\
70.96 &384.8&45648.7& 8239.8&0.0368 &0.0644\\
89.34 &734.2&88337.9& 16036.1& 0.0101& 0.0382\\
112.5 &1417.1& 168480.9& 30636.2& 0.0194& 0.0250\\
141.6 &2565.8& 317727.8& 57230.3& $-$0.00396 &0.0219\\
178.3 &4821.6& 588892.8& 106083.7& 0.0101 &0.0134\\
224.4 &8631.8& 1070807.1& 192603.7& $-$0.00296& 0.00672\\
282.5 &15376.1&1912774.1& 342706.1& 0.00214& 0.00953\\
\enddata
\tablecomments{\footnotesize
  Result for all fields. ${\rm DD}_{\rm mean}$, ${\rm RR}_{\rm mean}$ and ${\rm DR}_{\rm mean}$ are the mean numbers of quasar-quasar
  , random-random and quasar-random pairs within each $s$ bin for the ten jackknife samples. $\xi(s)$ is the mean value calculated
  from jackknife samples, and the error quoted is that from the
  jackknifes as well.
}
\end{deluxetable}

\begin{figure*}
  \centering
    \includegraphics[width=0.45\textwidth]{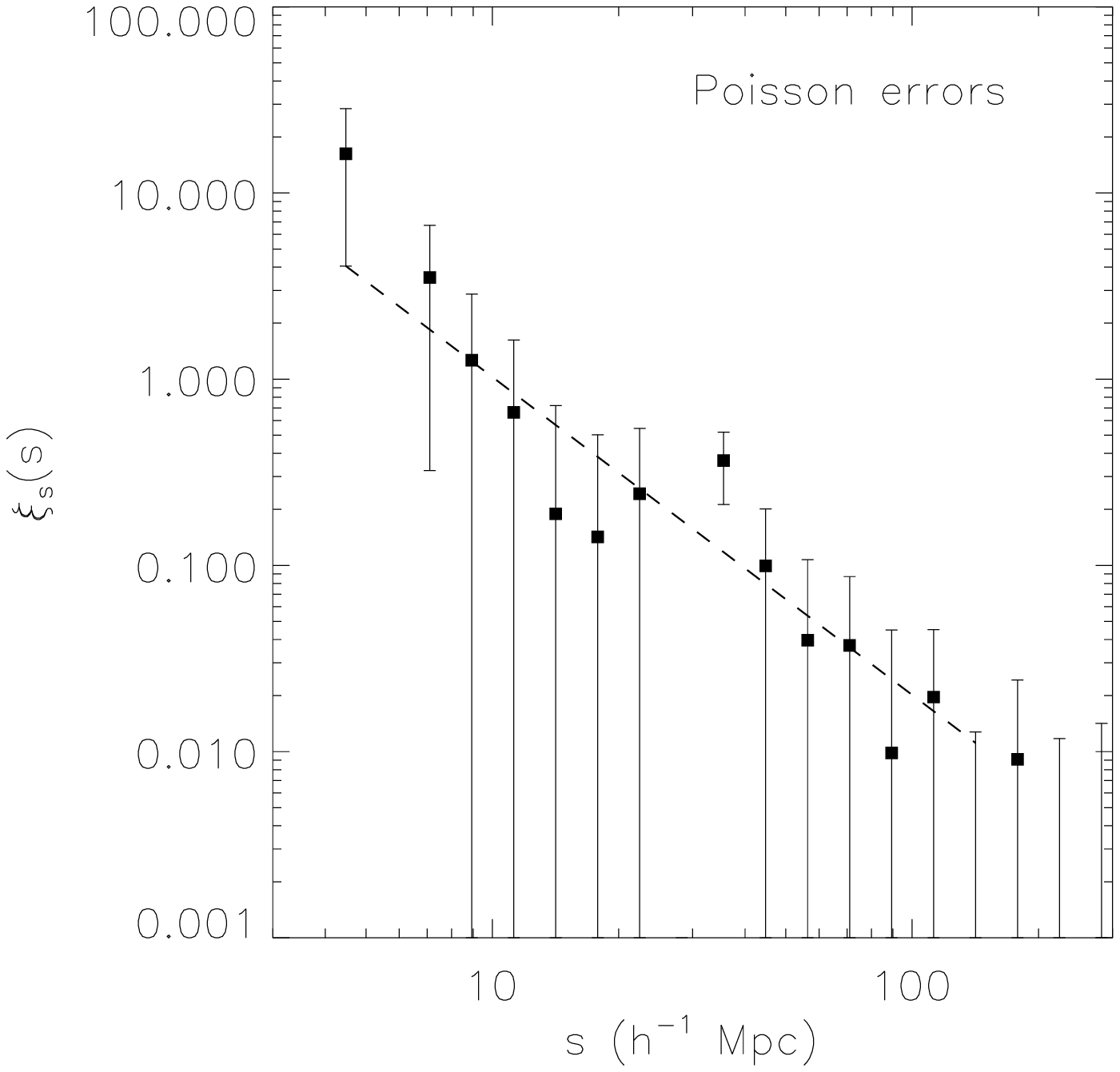}
    \includegraphics[width=0.45\textwidth]{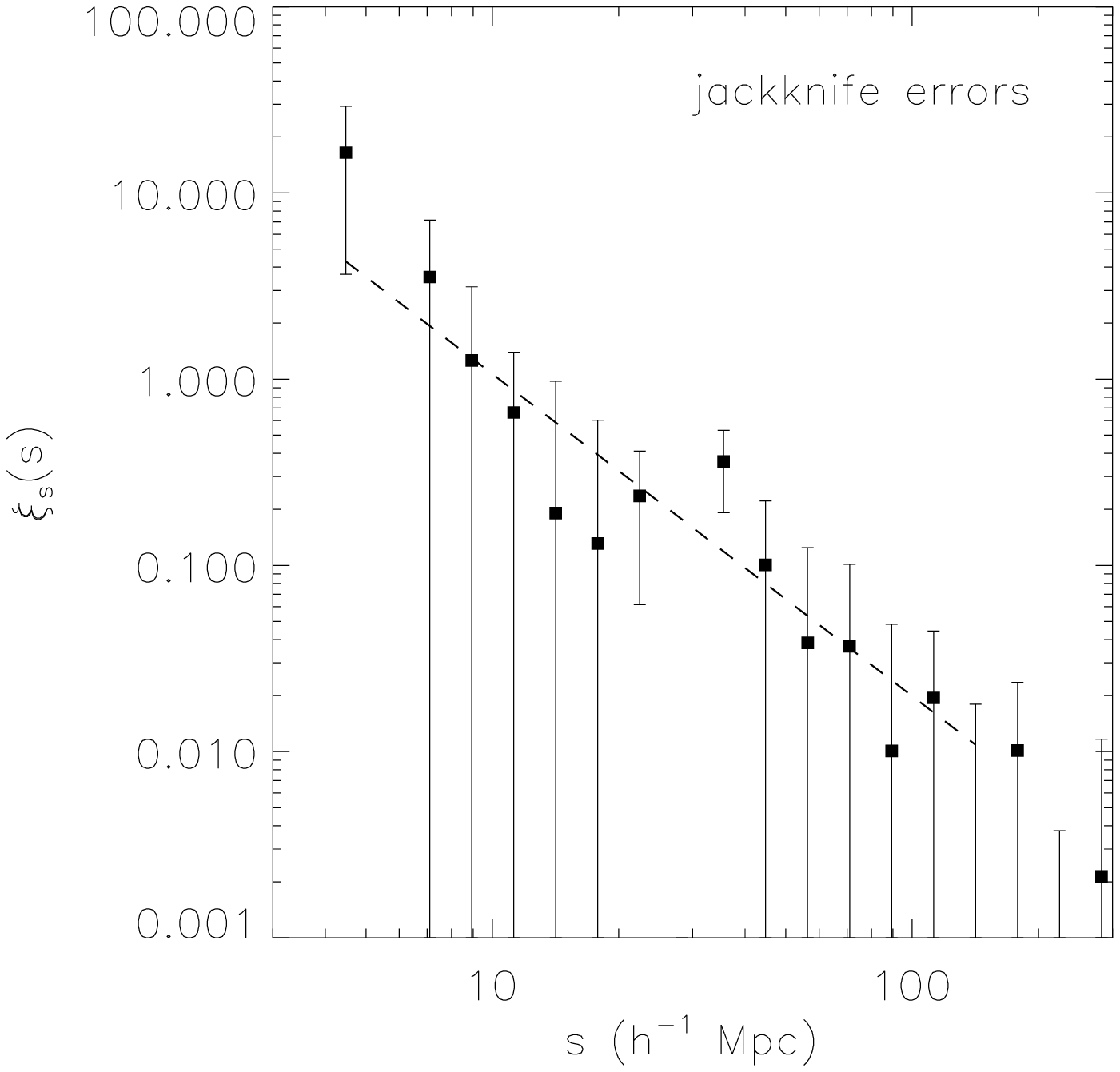}
    \caption{Redshift space correlation function $\xi_s(s)$ for quasars with $z\ge 2.9$ (all fields included). Statistical errors are
    estimated using the Poisson estimator ({\em left}) and jackknife estimator ({\em right}).
    The two estimators give comparable results. Also plotted are the best fitted power-law functions,
    with fitted parameters listed in Table \ref{table:fits}.}
    \label{fig:xi_s}
\end{figure*}

We draw random quasar catalogs according to the detailed angular and
radial selection functions discussed in the last section.

We start by computing the correlation function in ``redshift
space'', where each object is placed at the comoving distance
implied by its measured redshift and our assumed cosmology, with
no correction for peculiar velocities or redshift
errors\footnote{All calculations in this paper are done in
comoving coordinates, which is appropriate for comparing
clustering results at different epochs on linear scales. On very
small, virialized scales,  Hennawi \etal\ (2006a) argue that
proper coordinates are more appropriate for clustering analyses.}.
The correlation function is measured using the estimator of Landy
\& Szalay (1993)\footnote{We found that the Hamilton (1993)
estimator gives similar results.}:
\begin{equation}
\xi_s(s)=\frac{\bracket{DD}-2\bracket{DR}+\bracket{RR}}{\bracket{RR}}\ ,
\label{eq:landy_szalay}
\end{equation}
where $\bracket{DD}$, $\bracket{DR}$, and $\bracket{RR}$ are the
normalized numbers of data-data, data-random and random-random
pairs in each separation bin, respectively. The results are shown
in Fig.~\ref{fig:xi_s}, where we bin the redshift space distance
$s$ in logarithmic intervals of $\Delta \log_{10} s = 0.1$.  We
tabulate the results in Table~\ref{table:xi_s}.

There are various ways to estimate the statistical errors in the
correlation function (e.g., Hamilton 1993), including bootstrap
resampling (e.g., PMN04), jackknife resampling (e.g., Zehavi
\etal\ 2005), and the Poisson estimator (e.g., Croom \etal\ 2005;
da \^{A}ngela \etal\ 2005). In this paper we will focus on the
latter two methods. For the jackknife method, we split the
clustering sample into 10 spatially contiguous subsamples, and our
jackknife samples are created by omitting each of these subsamples
in turn. Therefore, each of the jackknife samples contains 90\% of
the quasars, and we use each to compute the correlation function.
The covariance error matrix is estimated as
\begin{equation}
{\rm Cov}(\xi_i,
\xi_j)=\frac{N-1}{N}\sum_{l=1}^N\left(\xi_i^l-\bar{\xi}_i\right)\left(\xi_j^l-\bar{\xi}_j\right)\
,
\end{equation}
where $N=10$ in our case, the subscript denotes the bin number,
and $\bar{\xi}_i$ is the mean value of the statistic $\xi_i$ over
the jackknife samples (not surprisingly, we found that
$\bar{\xi}_i$ was very close to the correlation function for the
whole clustering sample, for all bins $i$). Our sample is sparse,
thus the off-diagonal elements of the covariance matrix
are poorly determined, so we use only the diagonal
elements of the covariance matrix in the $\chi^2$ fits below.  We
also carried out fits keeping those off-diagonal elements for
adjacent and separated-by-two bins, and found similar results.

\begin{figure*}
  \centering
    \includegraphics[width=\textwidth]{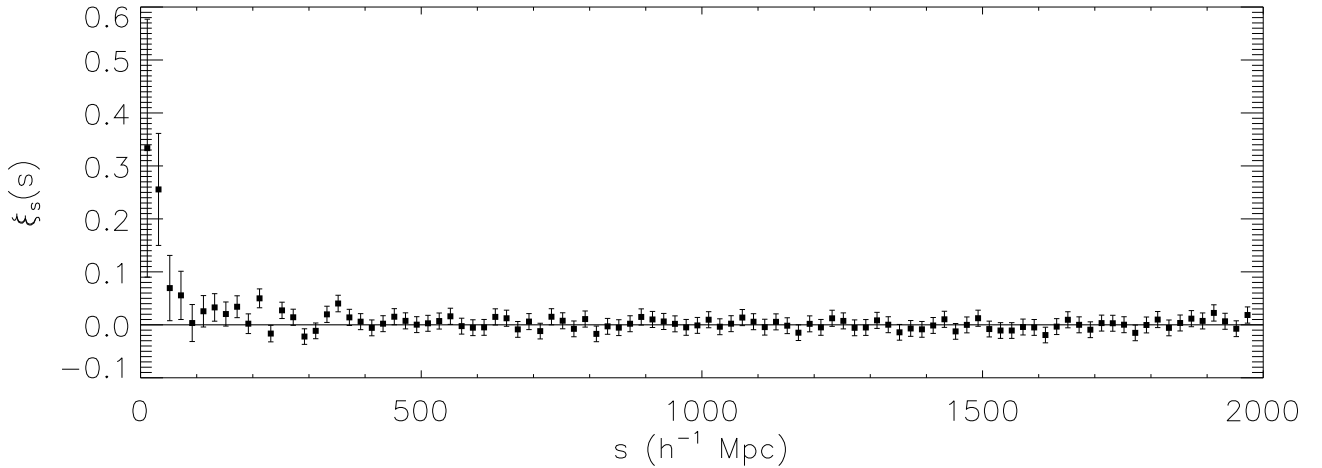}
    \caption{Large scale behavior of $\xi_s(s)$ for the $z\ge 2.9$ quasars (all fields included). Errors are estimated using
    the Poisson estimator. The redshift space correlation function
    essentially vanishes after $s>200\ h^{-1}$ Mpc, with a mean of 0.002 and rms scatter $\pm 0.01$
    in the range $200<s<2000\ h^{-1}$ Mpc.}
    \label{fig:xi_s_ls_linear}
\end{figure*}

For the Poisson error estimator (e.g., Kaiser 1986), valid for
sparse samples in which a given quasar is unlikely to take part in
more than one pair, the error is estimated as
$\Delta\xi_i=(1+\xi_i)/\sqrt{{\rm Min}(N_{\rm pair},N_{\rm
QSO})}$, where $N_{\rm pair}$ is the number of unique
quasar-quasar pairs in our real quasar sample in the bin in question,
and $N_{\rm QSO}$ is
the total number of real quasars in our sample (e.g., da
\^{A}ngela \etal\ 2005). The Poisson estimator breaks down on
large scales, as the pairs in different bins become correlated.
Fig.~\ref{fig:xi_s} shows the two error estimators; the two
methods give similar results.

The correlation function lies above unity for scales below $\sim
10\ h^{-1}$ Mpc; it is clear that the clustering signal is much
stronger than that of low-redshift quasars (e.g., Croom \etal\
2005; Connolly \etal\ 2006). Fig.~\ref{fig:xi_s} also shows the
results of a $\chi^2$ fit of a power-law correlation function
$\xi_s(s)=(s/s_0)^{-\delta}$ to the data with $4 < s < 150\
h^{-1}$ Mpc. The clustering signal is negative in the
$s=28.25{\rm\ h^{-1}Mpc}$ bin; Table \ref{table:xi_s} shows a smaller
number of quasar-quasar pairs than expected.  This point appears to be
an outlier, as the expected correlation function should be positive on
these scales; this discrepancy may be due to the paucity of quasars in
the sample at $z \sim 3.5$.  We have carried out fits to $\xi_s(s)$ both
including and not including this data point (Table~\ref{table:fits});
we find it makes little difference.
In particular, neglecting the point at 28.25 $h^{-1}$ Mpc, we find
$s_0=10.2\pm 3.1\ h^{-1}$ Mpc and $\delta=1.71\pm 0.43$ for the
Poisson errors, and $s_0=10.4\pm 3.0\ h^{-1}$ Mpc and
$\delta=1.73\pm 0.46$ for the jackknife method.  When we include
this negative data point, we find $s_0=10.4\ h^{-1}$Mpc and
$\delta=2.07$ for the jackknife method.
  Table~\ref{table:fits} also
includes the $\chi^2/\rm dof$ for these fits; in all cases, it is less
than unity, due to our neglecting the off-diagonal elements in the
covariance matrix.  However, as Figure~\ref{fig:xi_s} makes clear, the
majority of the points lie within 1 sigma of the fitted power law.

Using good fields only yields similar results for bins where there
are more than 20 real quasar pairs (i.e., $s\gtrsim 20\ h^{-1}$
Mpc).
On scales below 20 $h^{-1}$ Mpc there are very few quasar-quasar
pairs in each bin, and the signal-to-noise ratio is very low.
The fitting results (over scale range $4 < s < 150\ h^{-1}$ Mpc)
are: $s_0=12.7\pm 3.3\ h^{-1}$ Mpc and $\delta=1.64\pm 0.31$
for the Poisson errors; $s_0=10.3\pm3.0\ h^{-1}$ Mpc and
$\delta=1.43\pm 0.28$
for the jackknife errors.

To study the large scale behavior of $\xi_s(s)$ we compute
$\xi_s(s)$ up to $s= 2000$ $h^{-1}$ Mpc on a linear grid with
$\Delta s=20\ h^{-1}$ Mpc, using all the fields. The result is
shown in Fig.~\ref{fig:xi_s_ls_linear} and errors are estimated
using the Poisson estimator. For scales $200<s<2000\ h^{-1}\,$Mpc,
the mean value of $\xi_s(s)$ is 0.002, with an rms scatter of $\pm
0.01$ (see also Roukema, Mamon \& Bajtlik 2002 and Croom \etal\
2005). Thus there is no clear evidence for correlations on scales
above $200 \ h^{-1}$ Mpc.

\subsection{The Real Space Correlation Function}
\label{sec:real_space}

\begin{figure*}
  \centering
    \includegraphics[width=0.45\textwidth]{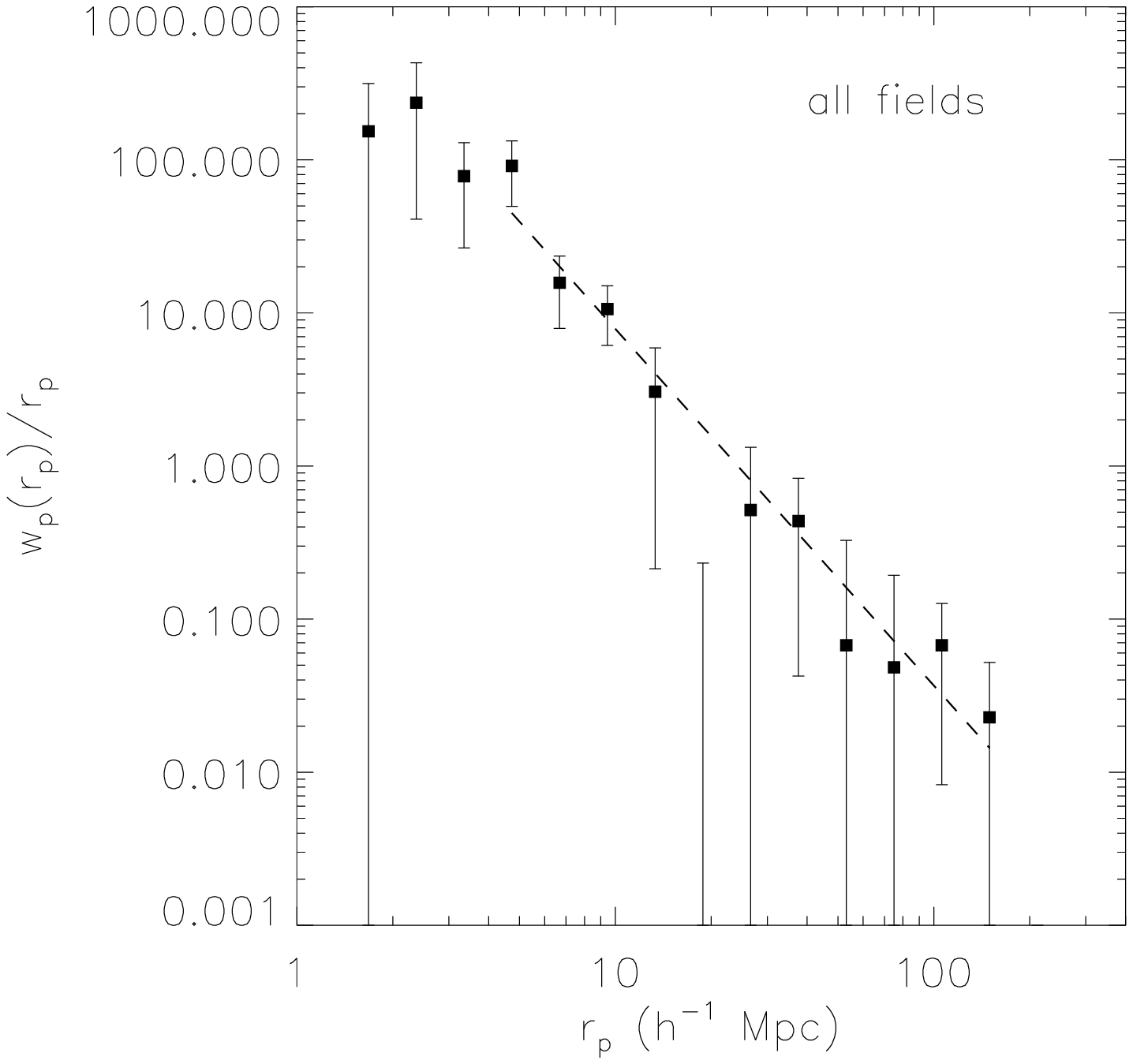}
    \includegraphics[width=0.45\textwidth]{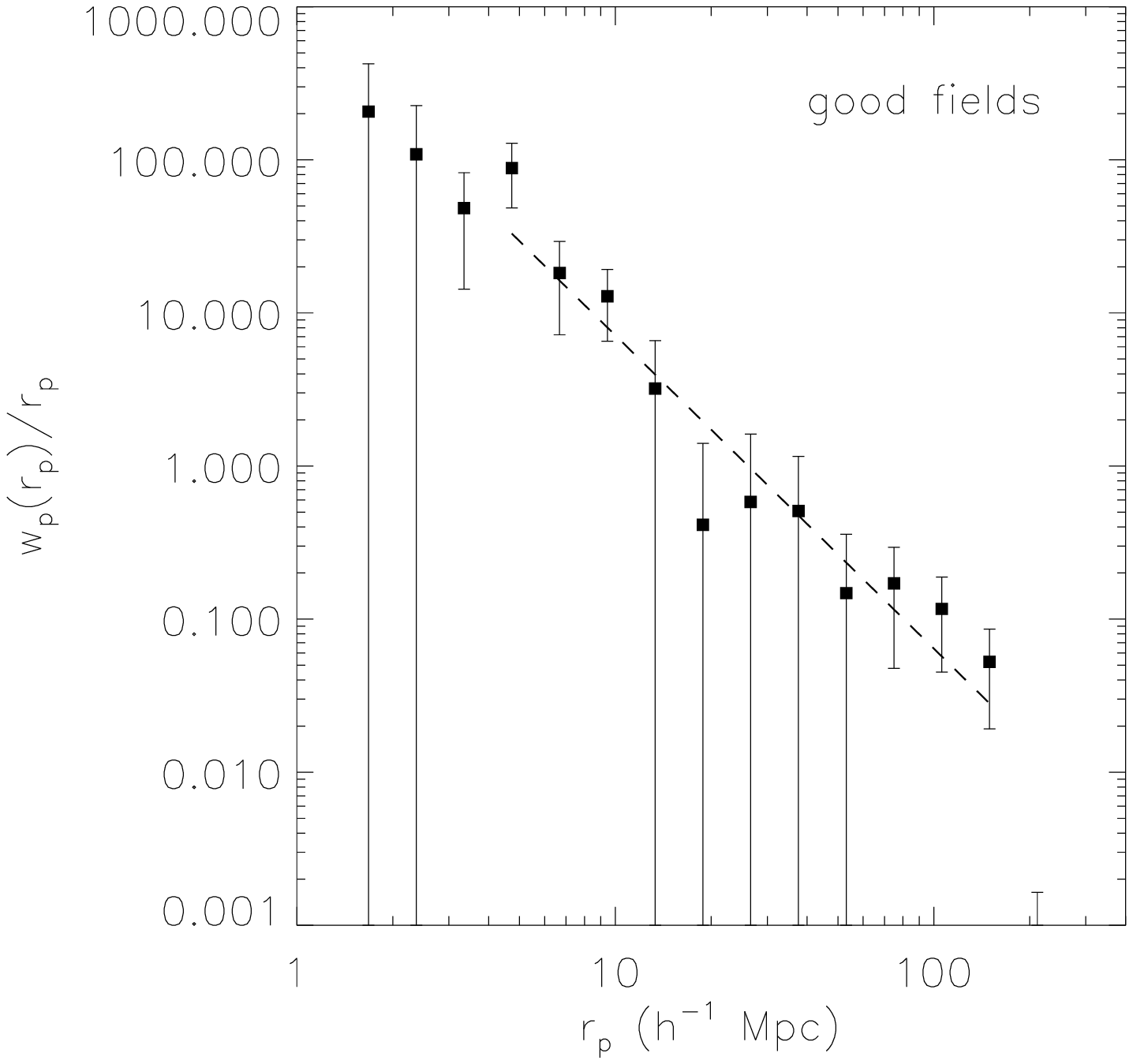}
    \caption{Projected correlation function $w_p(r_p)$ for the $z\ge 2.9$ quasars. Errors are estimated using the jackknife method. Also
    plotted are the best fitted power-law functions, with fitted
    parameters listed in Table \ref{table:fits}. {\em left}: for all fields;
    {\em right}: for good fields only. The two cases give
    similar results.}
    \label{wp}
\end{figure*}

\begin{deluxetable}{lrrrrr}
\tablecolumns{6} \tablewidth{0.45\textwidth}
\tablecaption{Projected correlation function
$w_p(r_p)$\label{table:w_p}} \tablehead{$r_p$ ($h^{-1}$ Mpc) &
${\rm DD}_{\rm mean}$ &${\rm RR}_{\rm mean}$ & ${\rm DR}_{\rm
mean}$ & $\frac{w_p}{r_p}$ & $\frac{w_p}{r_p}$ error} \startdata
1.189 &0.0      &114.3     &19.8     &--      & --   \\
1.679 &0.9      &258.3     &39.6     &154     & 162  \\
2.371 &4.5      &478.5     &91.8     &236     & 195  \\
3.350 &9.9      &913.2     &160.8    &78.1    & 51.5 \\
4.732 &20.7     &1864.1    &359.9    &91.3    & 41.6 \\
6.683 &32.4     &3786.5    &684.3    &15.7    & 7.81 \\
9.441 &62.9     &7158.5    &1314.0   &10.6    & 4.45 \\
13.34 &130.0    &14551.2   &2659.1   &3.06    & 2.85 \\
18.84 &227.3    &28598.1   &5162.4   &-0.681  &0.913 \\
26.61 &488.5    &56940.7   &10123.8  &0.516   &0.810 \\
37.58 &871.7    &111284.0  &19955.6  &0.437   &0.395 \\
53.09 &1762.2   &218346.8  &38910.9  &0.0675  &0.259 \\
74.99 &3394.4   &422580.9  &75630.1  &0.0484  &0.145 \\
105.9 &6751.7   &811406.0  &145785.5 &0.0674  &0.0592 \\
149.6 &12425.7  &1535320.8 &274851.9 &0.0228  &0.0292 \\
211.3 &22655.1  &2849970.6 &509877.9 &-0.0183 &0.00992 \\
\enddata
\tablecomments{\footnotesize
  Result for all fields. ${\rm DD}_{\rm mean}$, ${\rm DR}_{\rm mean}$,
 and ${\rm RR}_{\rm mean}$ are the mean
  numbers of quasar-quasar, random-random and quasar-random
  pairs within each $r_p$ bin for the ten jackknife samples. $w_p(r_p)/r_p$ is the mean value calculated
  from the jackknife samples.}
\end{deluxetable}

\begin{figure*}
  \centering
    \includegraphics[width=0.45\textwidth]{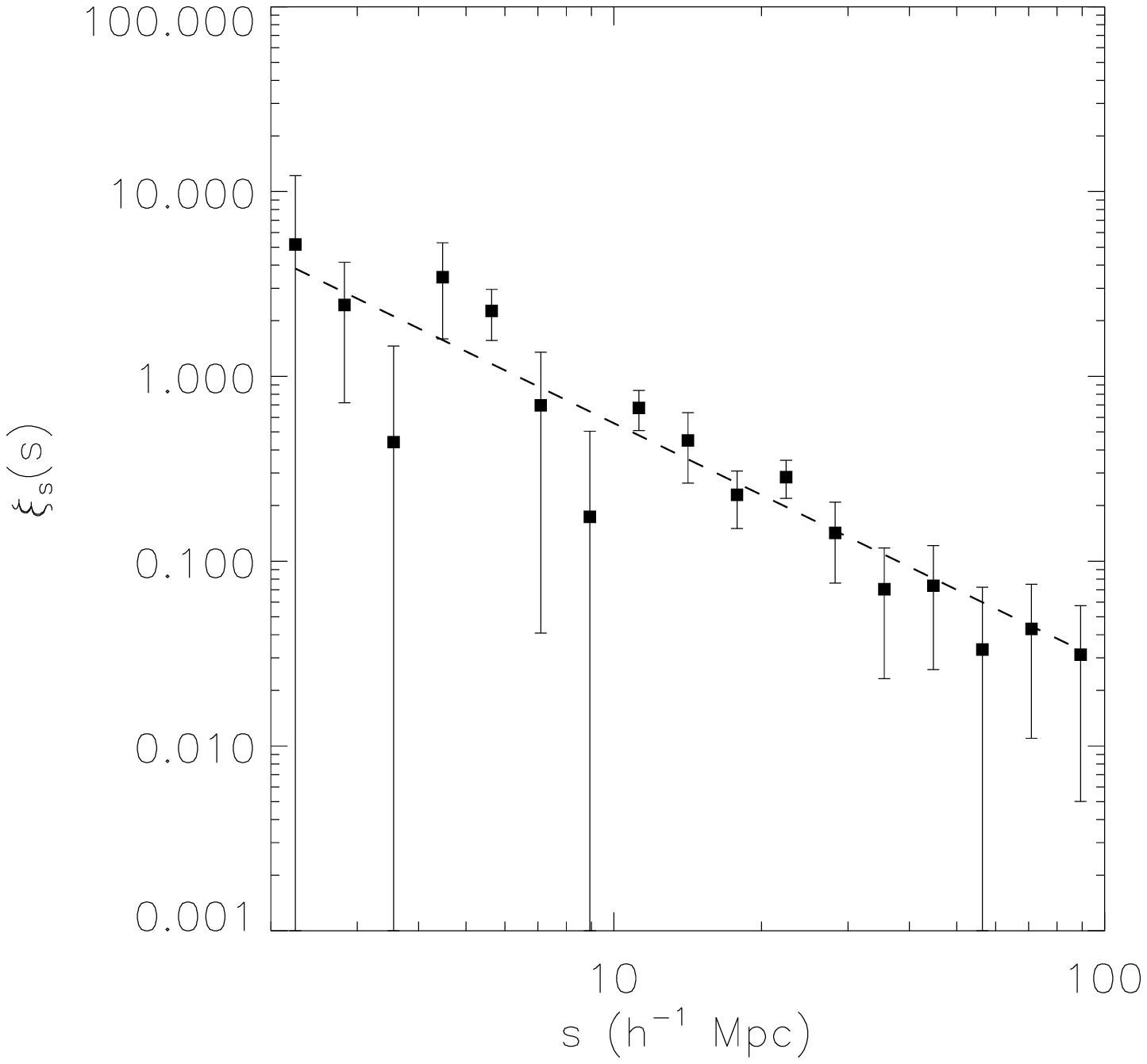}
    \includegraphics[width=0.45\textwidth]{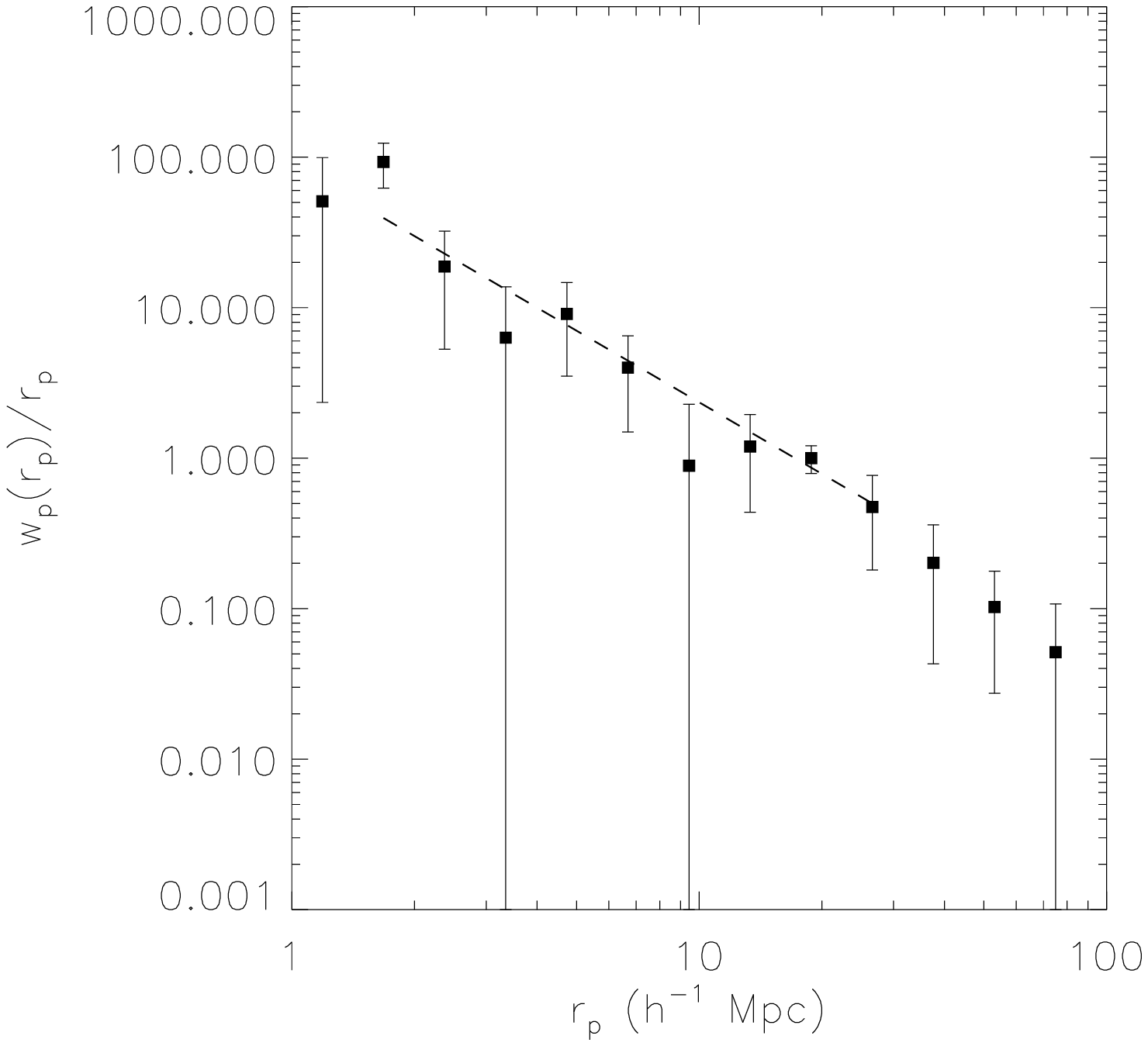}
    \caption{Correlation functions of 23,283 $0.8\le z\le 2.1$ SDSS
      DR5 quasars in all fields.
    Errors are estimated using the jackknife method. {\em left}: redshift space correlation function;
    {\em right}: projected correlation function. Also
    plotted are the best fitted power-law functions, with fitted
    parameters listed in Table \ref{table:fits}.}
    \label{fig:lowz_wp}
\end{figure*}

\begin{figure*}
  \centering
    \includegraphics[width=0.45\textwidth]{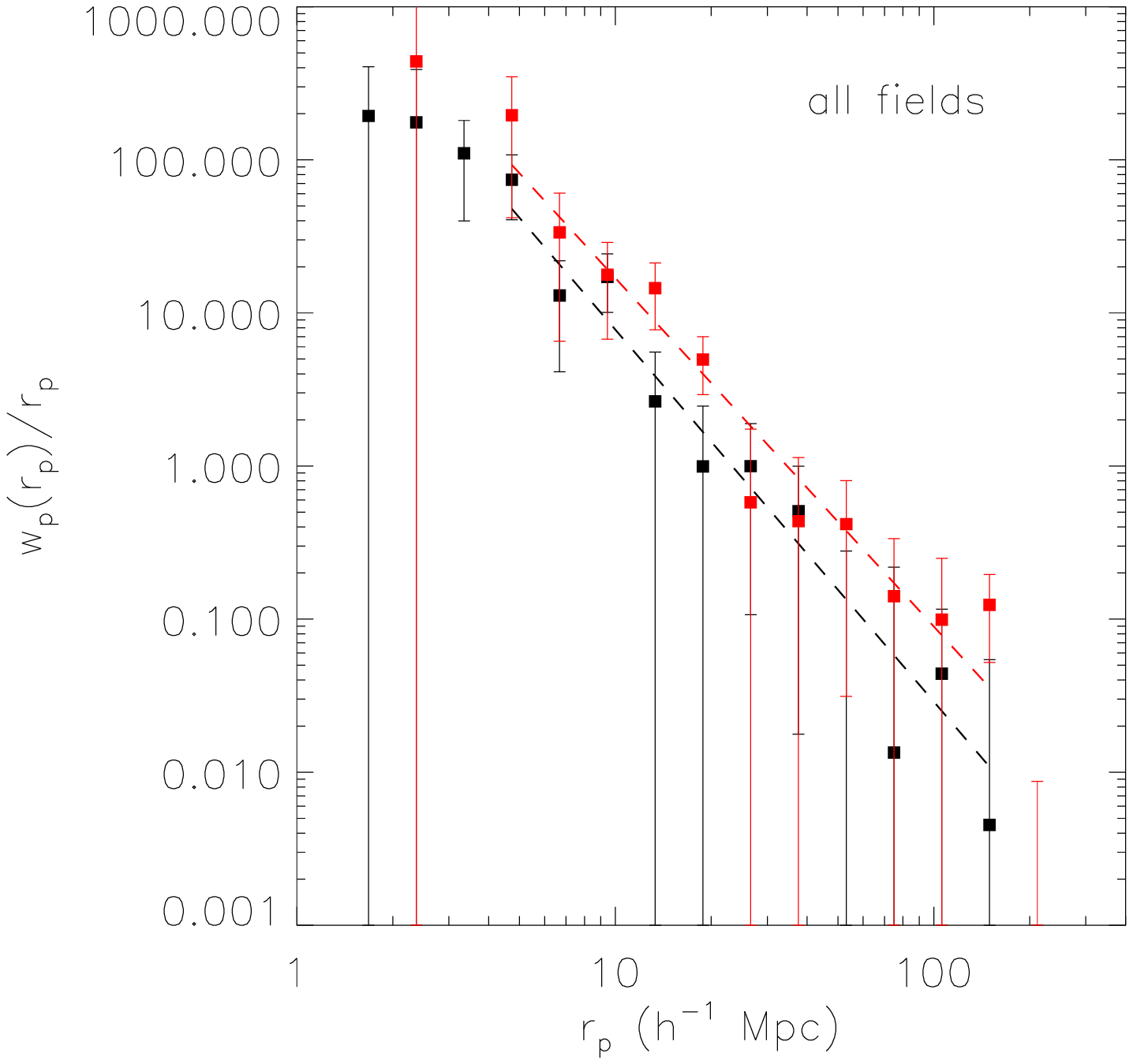}
    \includegraphics[width=0.45\textwidth]{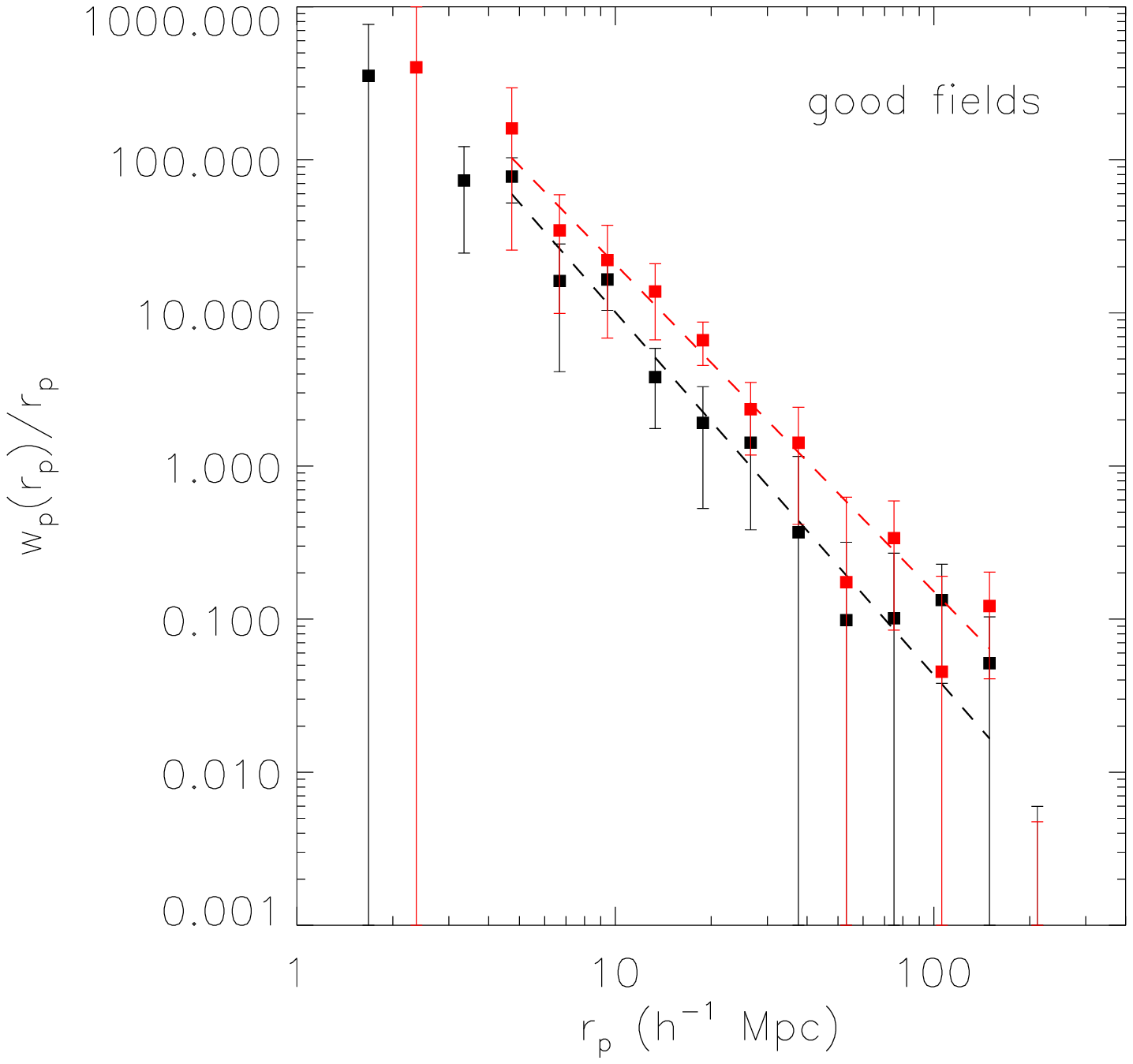}
    \caption{Clustering evolution of high redshift quasars.
    Errors are estimated using the jackknife method.
    Black indicates the $2.9\le z\le 3.5$ bin and red indicates the $z\ge 3.5$ bin. Also
    plotted are the best fitted power-law functions, with fitted
    parameters listed in Table \ref{table:fits}. {\em left}: all fields; {\em right}: good fields
    only. Both cases show stronger clustering in the
    higher redshift bin.}
    \label{fig:wp_z_evo}
\end{figure*}

Appendix~\ref{app:redshift} shows that the uncertainty in
measurements of the quasar redshifts is substantial, $\Delta z
\approx 0.01$, giving an uncertainty in the comoving distance of a
$z=3.5$ quasar of $\sim 6\ h^{-1}$ Mpc. This, together with
peculiar velocities on large and small scales systematically bias
the correlation function (e.g., Kaiser 1987). To determine the
real-space correlation function, we follow standard practice and
compute the correlation function on a two-dimensional grid of pair
separations parallel ($\pi$) and perpendicular ($r_p$) to the line
of sight. Our grid has a logarithmic increment of 0.15 along the
$r_p$ direction and a linear increment of $5\ h^{-1}$ Mpc along
the $\pi$ direction. As above, the two dimensional correlation
function $\xi_s(r_p,\pi)$ is estimated using the Landy \& Szalay
(1993) estimator, equation~(\ref{eq:landy_szalay}). Redshift
errors and peculiar velocities affect the separation along the
$\pi$ direction but not along the $r_p$ direction. Therefore we
project out these effects by integrating $\xi_s(r_p,\pi)$ along
the $\pi$ direction to obtain the projected correlation function
$w_p(r_p)$:
\begin{equation}
w_p(r_p)=2\int_0^\infty d\pi\,\xi_s(r_p,\pi)\ .
\end{equation}
In practice we integrate up to some cutoff value of $\pi_{\rm
cutoff} = 100\ h^{-1}$ Mpc, which includes most of the clustering
signal, without being dominated by noise. This value of $\pi_{\rm
cutoff}$ is larger than the values of $40-70\ h^{-1}$ Mpc
typically used in clustering analyses for galaxies and
low-redshift quasars (e.g., Zehavi \etal\ 2005, PMN04, da
\^{A}ngela \etal\ 2005) because of the substantially stronger
clustering of high-redshift quasars. We verify that our results
are not sensitive to the precise value of $\pi_{\rm cutoff}$ we
adopt.

The projected correlation function $w_p$ is related to the
real-space correlation function $\xi(r)$ through
\begin{equation}
w_p(r_p)=2\int_{r_p}^\infty\frac{r\xi(r)}{(r^2-r_p^2)^{1/2}}dr\
\end{equation}
(e.g., Davis \& Peebles 1983).

If $\xi(r)$ follows the power-law form
$\xi(r)=(r/r_0)^{-\gamma}$, then:
\begin{equation}
\frac{w_p(r_p)}{r_p}=\frac{\Gamma(1/2)\Gamma[(\gamma-1)/2]}{\Gamma(\gamma/2)}\left(\frac{r_0}{r_p}\right)^{\gamma}\
.
\end{equation}

\begin{deluxetable*}{lcccccccc}
%\rotate
\tablecolumns{8} \tablewidth{\textwidth} \tablecaption{Summary of
the fitting parameters of the correlation function
\label{table:fits}} \tablehead{redshift & case&$\xi_s(s)/\xi(r)$ &
$s_0/r_0$ ($h^{-1}$ Mpc) &$\delta/\gamma$ & $\chi^2/$dof
&$s_0/r_0\ (\delta,\, \gamma=2.0)$ & $\chi^2$/dof} \startdata
$z\ge 2.9$ & all, Poisson  &$(s/s_0)^{-\delta}$   & $10.16\pm 3.08$  & $1.71\pm 0.43$ & 0.47 \\
           & all, jackknife&$(s/s_0)^{-\delta}$  & $10.39\pm 3.00$  & $1.73\pm 0.46$  & 0.37\\
           & all, jackknife\tablenotemark{a}& $(s/s_0)^{-\delta}$  & $10.38\pm 2.57$  & $2.07\pm 0.62$ & 0.62 \\
           & good, Poisson &$(s/s_0)^{-\delta}$   & $12.72\pm 3.25$  & $1.64\pm 0.31$ & 0.35 \\
           & good, jackknife&$(s/s_0)^{-\delta}$ & $10.28\pm 2.95$  & $1.43\pm 0.28$ & 0.46\\
\\
$z\ge 2.9$ & all, jackknife &$(r/r_0)^{-\gamma}$  & $16.10\pm 1.70$  & $2.33\pm 0.32$ & 0.32 &$14.71\pm 1.86$ & 0.42\\
           &all, jackknife\tablenotemark{a}&$(r/r_0)^{-\gamma}$  & $13.60\pm 1.83$  & $3.52\pm 0.87$ & 0.75\\
           & good, jackknife &$(r/r_0)^{-\gamma}$  & $15.16\pm 2.75$  & $2.05\pm 0.28$ &0.75 &$14.81\pm 1.94$ & 0.68\\
\\
$2.9\le z\le3.5$ & all, jackknife  &$(r/r_0)^{-\gamma}$   & $16.02\pm 1.81$ & $2.43\pm 0.43$ &0.43 &$14.79\pm 2.12$ & 0.52\\
                 & good, jackknife  &$(r/r_0)^{-\gamma}$  & $17.91\pm 1.51$ & $2.37\pm 0.29$ &0.46 & $16.90\pm 1.73$ & 0.56\\
\\
$z\ge 3.5$  & all, jackknife  &$(r/r_0)^{-\gamma}$ & $22.51\pm 2.53$  & $2.28\pm 0.31$ & 0.50 &$20.68\pm 2.52$ & 0.52\\
            & good, jackknife &$(r/r_0)^{-\gamma}$   & $25.22\pm 2.50$  & $2.14\pm 0.24$ & 0.32 &$24.30\pm 2.36$ & 0.32\\
$0.8\le z\le 2.1$ & all, jackknife  &$(s/s_0)^{-\delta}$   & $6.36\pm 0.89$  & $1.29\pm 0.14$ & 0.88 \\
                  & all, jackknife &$(r/r_0)^{-\gamma}$   & $6.47\pm 1.55$  & $1.58\pm 0.20$ & 0.88 \\
\enddata
\tablenotetext{a}{Data points with negative correlation function
are included in the fit.} \tablecomments{\footnotesize Fitting
results for various cases and different redshift bins. The {\em
case} column indicates whether the correlation function is
measured from all fields or from good fields only; it also
indicates the error estimator. $\xi_s(s)$ is the redshift space
correlation function, while $\xi(r)$ is the real space correlation
function. The last two columns give the correlation length and
reduced $\chi^2$ for the fixed power-law index fits for selected
cases.}
\end{deluxetable*}

We show our results for $w_p(r_p)$ in Fig.~\ref{wp}, where the
errors are estimated using the jackknife method. Tabulated values
for $w_p$ are listed in Table \ref{table:w_p} for the all-fields
case. We only use data
points
where the mean number of quasar-quasar pairs in the $r_p$ bin is
more than 10, and we therefore restrict our fits to scales $4
\lesssim r_p\lesssim 150\ h^{-1}$ Mpc. The parameters of the
best-fit power-law for the all-fields case is $r_0=16.1\pm 1.7\
h^{-1}$ Mpc and $\gamma=2.33\pm 0.32$ when the negative data point
at $r_p=18.84\ h^{-1}$ Mpc is excluded. When this negative data
point is included in the fit we get $r_0=13.6\pm 1.8\ h^{-1}$ Mpc
and an unusually large $\gamma=3.52\pm 0.87$, which is caused by
the drag of the negative point on the fit\footnote{For the
good-fields case the projected correlation function is positive
over the full range that we fit.}.
Using good fields only yields $r_0=15.2\pm 2.7\ h^{-1}$ Mpc and
$\gamma=2.05\pm 0.28$,
shown in the lower panel of Fig.~\ref{wp}. Note that the
real-space correlation function indicates appreciably stronger
clustering than does its counterpart in redshift space; the large
redshift errors spread structures out in redshift space, diluting
the clustering signal.

We have already indicated that the clustering signal is
appreciably stronger than at lower redshift.  To check that this
was not somehow an artifact of our processing we selected a sample
of 23,283 spectroscopically confirmed quasars with $0.8 \le z \le
2.1$ from the SDSS DR5, with the same selection criteria as we
used above (\S~\ref{sec:cluster_sample}). Figure~\ref{fig:lowz_wp}
shows the resulting $\xi_s(s)$ and $w_p(r_p)$; to compare with the
results of other authors (e.g., da \~Angela \etal\ 2005; Connolly
\etal\ 2006), we integrated to $\pi_{\rm cutoff} = 70\ h^{-1}$
Mpc.  We fit power-laws over the range $1< s < 100\ h^{-1}$ Mpc
(Croom \etal\ 2005) for $\xi_s(s)$, and $1.2 < r_p < 30\ h^{-1}$
Mpc for $w_p(r_p)$ (PMN04 and da \^{A}ngela \etal\ 2005). The
fitted power-law parameters are: $s_0=6.36\pm 0.89$ $h^{-1}$ Mpc
and $\delta=1.29\pm 0.14$ for $\xi_s(s)$; $r_0=6.47\pm 1.55$
$h^{-1}$ Mpc and $\gamma=1.58\pm 0.20$ for
$w_p(r_p)$.
These results are in excellent agreement with Croom \etal\ (2005),
PMN04 and da \^{A}ngela \etal\ (2005) based on the 2QZ sample, and
Connolly \etal\ (2006) based on the SDSS sample.
Note that the 2QZ papers use a slightly different cosmology, which
causes very little difference. More importantly, the 2QZ sample is
at lower mean luminosity than the SDSS sample, although there is
only a mild luminosity dependence of the clustering strength (e.g.,
Lidz et al. 2006; Connolly et al. 2006).
We note that the amplitude of $w_p(r_p)$ for $r_p\gtrsim30$
$h^{-1}$ Mpc is lower than predicted from the power-law fit, which
is also the case in da \^{A}ngela \etal\ (2005, Fig.~2).

The predicted correlation function of the underlying dark
matter at $r = 15\ h^{-1}$ Mpc is $\sim 0.014$ at
$z=3.5$ (see \S\ref{subsec:qsolifetime} and Appendix
\ref{app:martini_weinberg}), far below that of the current high
redshift quasar sample (Fig.~\ref{wp}), indicating that our
high-redshift quasar sample is very strongly biased.

The increase in clustering signal with redshift we have seen
suggests that we may be able to see redshift evolution {\em
within}  our sample.  We divide our clustering sample into two
subsamples with redshift intervals $2.9\le z\le 3.5$ and $z\ge
3.5$. The resulting $w_p(r_p)$ are shown in
Fig.~\ref{fig:wp_z_evo}. The higher redshift bin shows
systematically stronger clustering than does the lower redshift
bin. The fitted parameters are: $r_0=16.0\pm 1.8$ $h^{-1}$ Mpc and
$\gamma=2.43\pm0.43$ for $2.9\le z \le 3.5$; and $r_0=22.5\pm 2.5$
$h^{-1}$ Mpc and $\gamma=2.28\pm0.31$ for $z\ge 3.5$, where the
fitting range is 4 $\sim$ 150 $h^{-1}$ Mpc. Using good fields only
yields: $r_0=17.9\pm 1.5$ $h^{-1}$ Mpc and $\gamma=2.37\pm0.29$
for $2.9\le z \le 3.5$; $r_0=25.2\pm 2.5$ $h^{-1}$ Mpc and
$\gamma=2.14\pm0.24$ for $z\ge 3.5$. When we fix the power-law
index to be $\gamma=2.0$ we get slightly different but consistent
correlation lengths for each case (Table \ref{table:fits}).
Indeed, the clustering of quasars increases strongly with redshift
over the range probed by our sample.

The increase in clustering strength with redshift may be due to
two effects: an ever-increasing bias of the halos hosting quasars
with fixed luminosity with redshift, and luminosity-dependent
clustering. The higher-redshift quasars are more luminous (Table
\ref{table:LF} and Fig.~17 of Richards \etal\ 2006), and may be
associated with more massive haloes.  At low redshift ($z\lesssim
3$) and moderate luminosities, luminosity depends on accretion
rate as much as black hole mass, and one expects little dependence
of clustering strength on luminosity (Lidz \etal\ 2006), as
observed (Croom \etal\ 2005; Connolly \etal\ 2006). However, the
high-luminosity high redshift quasars in our sample have close to
Eddington luminosities (Kollmeier \etal\ 2006), and therefore we
may well expect a strong dependence of the clustering signal on
luminosity (Hopkins \etal\ 2006).  We are limited by the
relatively small size of our sample to date, and will explore the
dependence of clustering strength with luminosity in a future
paper.

\begin{figure}
  \centering
    \includegraphics[width=0.48\textwidth]{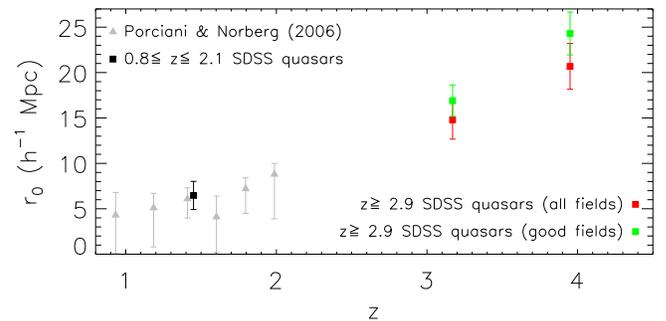}
    \caption{The evolution of the comoving correlation length $r_0$ as
    a function of redshift. Gray
    triangles are 2QZ data points taken from Porciani \& Norberg (2006, Column 7 in
    their table 3). The black
    square is for the $0.8\le z\le 2.1$ SDSS quasars, taken from the
    variable power-law index fit; the red and green squares are for
    the $2.9\le z\le 3.5$ bin and the $z\ge 3.5$ bin for the
    all fields for the good fields cases respectively, taken from the fixed
    $\gamma=2.0$ fits.}
    \label{fig:r0_z_evo}
\end{figure}

Figure~\ref{fig:r0_z_evo} shows the evolution of
comoving correlation length $r_0$ as a function of redshift, where the
data points for
low redshift bins (gray triangles) are taken from Porciani \&
Norberg (2006, the 2QZ sample). Data points for the SDSS quasar
sample in this paper are denoted as filled squares, placed at the
mean redshifts for each redshift bin. The black
square is for the $0.8\le z\le 2.1$ SDSS quasars, taken from the
variable power-law index fit; the red and green squares are for
the $2.9\le z\le 3.5$ bin and the $z\ge 3.5$ bin (with $\gamma$ fixed
to 2.0), both for the
all fields case and the good fields case.  There are many factors that affect the fitted value
of $r_0$: the 2QZ and the SDSS samples probe different luminosities, the
range of scales over which the power law is fit are different, and the
power-law indices $\gamma$ are different. Nevertheless, this figure
demonstrates that the clustering length of quasars increases
dramatically with redshift.

\subsection{Quasar Lifetime, Halo Mass, and Bias}\label{subsec:qsolifetime}

The clustering of quasars and their space density can be used to
constrain the average quasar lifetime $t_{\rm Q}$\footnote{Here we
define $t_{\rm Q}$ to be the total time that an accreting
supermassive black hole has a UV luminosity above the luminosity
threshold of our sample.  If the black hole is as old as its host
dark matter halo, then the duty cycle $t_{\rm Q}/t_{\rm H}$ is the
probability that we observe a quasar in this halo.  Indeed, while
the equations in Appendix~\ref{app:martini_weinberg} show that the
directly constrained quantity is the duty cycle, the quantity
$t_{\rm Q}$ indicates how much time a supermassive black hole
spends during the luminous accretion phase as it assembles most of
its mass.} and the bias of the dark matter halos in which they sit
(Martini \& Weinberg 2001; Haiman \& Hui 2001). In this section,
we follow Martini \& Weinberg (2001); the essential formulas are
presented in Appendix~\ref{app:martini_weinberg}. The basic
assumptions are that: 1) luminous quasars only reside in dark
matter halos with mass above some threshold mass $M_{\rm min}$; 2)
those dark matter halos with $M\ge M_{\rm min}$ host at most one
active quasar at a time.  The probability that such a halo harbors
an active quasar is the duty cycle $t_{\rm Q}/t_{\rm H}$, where
$t_{\rm H}$ is the halo lifetime, given by eqn.~(\ref{t_halo}).
Assumptions (1) and (2) include the assumption that every dark
matter halo harbors a supermassive black hole, either active or
dormant, and that the resulting quasars have the
same clustering strength as their hosting halos.

We note that the Martini \& Weinberg approach is appropriate for
high redshift quasars because at low redshift ($z<2$), the
occurrence of quasar activity is determined by fuelling as well,
rather than by the mere existence of a dark matter halo. Therefore
the
probability that a halo harbors an active quasar is the duty cycle
$t_{\rm Q}/t_{\rm H}$ times the (unknown)
probability that a halo harbors an active or dead quasar.

\begin{figure}
  \centering
  \includegraphics[width=0.5\textwidth]{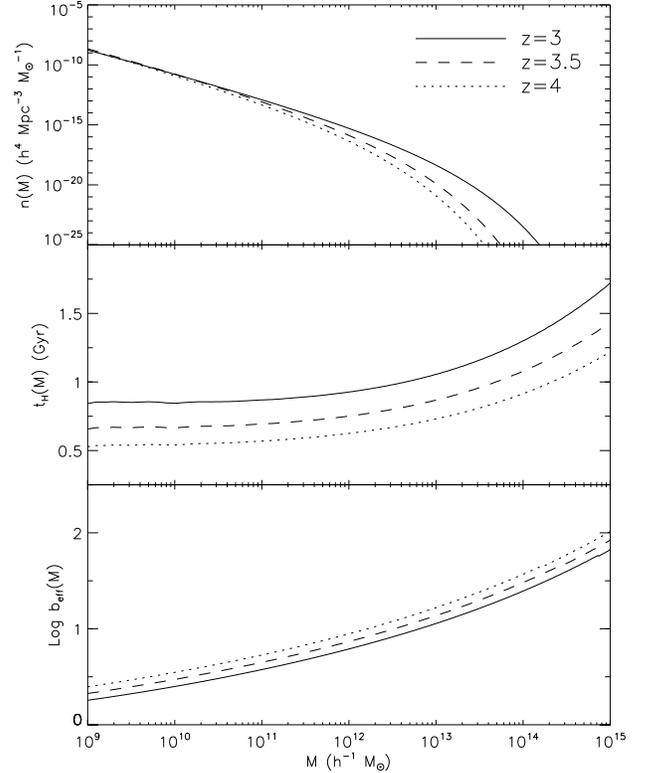}
  \caption{The Sheth \& Tormen (1999) halo mass function, halo lifetime and effective bias factors for
  halos with $M>M_{\rm min}$ as functions of halo mass for three redshifts $z=3,\ 3.5,\ 4$, in our
  fiducial cosmology.  The age of the universe at these three
  redshifts is 2.2, 1.9, and 1.6 Gyr, respectively, and for typical
  halos with a mass of a few $\times 10^{12}\ h^{-1}M_\odot$, the halo lifetime is approximately
  $0.7\sim 1$ Gyr at these redshifts.}
  \label{fig:nm}
\end{figure}

The value of $M_{\rm min}(z)$  is related to the quasar lifetime
and the observed quasar spatial density $\Phi(z)$ integrated over
the survey magnitude range (having corrected for the selection
function, of course):
\begin{equation}\label{eq:Phi_z}
\Phi(z)=\int_{M_{\rm min}}^\infty dM\frac{t_{\rm Q}}{t_{\rm
H}(M,z)}n(M,z)\ ,
\end{equation}
where we set the duty cycle $t_{\rm Q}/t_{\rm H}$ equal to unity
in the integration when $t_{\rm Q}>t_{\rm H}$, and $n(M,z)$ is the
dark matter halo mass function. Here, we follow Sheth \& Torman
(1999) to compute $n(M,z)$.
Given $\Phi(z)$ and assumed constant $t_{\rm Q}$, we can determine
$M_{\rm min}(z)$ from equation (\ref{eq:Phi_z}) and hence the
effective bias $b_{\rm eff}(M_{\rm min},z)$ from equation
(\ref{b_eff}), for which we have used the analytical bias
formalism in Jing (1998). We have checked the accuracy of the
analytical bias model using the results of a cosmological N-body
simulation by Paul Bode and Jeremiah P. Ostriker (Bode 2006,
private communication). At the simulation output redshifts, $z=3$ and
$z=4$, the bias factor depends on scale.  However, we will integrate
over a range of scales (see Eq.~\ref{xi_20} below), the scale-independent analytical bias
formalism provides an adequate prescription (see further discussion
in Appendix \ref{app:martini_weinberg}). More importantly, the
analytic form allows us to interpolate the bias with redshift, which
is needed to predict the observed correlation function
(equation~\ref{xi_average}). Fig.~\ref{fig:nm} shows $n(M,z)$, $t_{\rm
H}(M,z)$ and $b_{\rm eff}(M,z)$ as functions of halo mass $M$ (in
units of $h^{-1}\ M_\odot$) at redshift $z=3,\ 3.5$, and 4 for our
standard cosmology.

We compute the model predicted quasar correlation function
$\xi_{\rm model}(r,z)=b_{\rm eff}^2\xi_m(r,z)$ in steps of 0.1 in
redshift, and integrate it to
obtain the averaged correlation function $\bar{\xi}(r)$ over some
redshift range via equation (\ref{xi_average}).
$\bar{\xi}(r)$ is to be compared with our
measured correlation function $\xi(r)$. We iterate until we find a
proper $t_{\rm Q}$ to minimize the difference between $\xi(r)$ and
$\bar{\xi}(r)$. In practice, to compare the data and the model, we
use the {\em integrated correlation function} within $[r_{\rm
min},\ r_{\rm max}]\ h^{-1}$ Mpc, defined as
\begin{equation}\label{xi_20}
\xi_{20}=\frac{3}{r_{\rm max}^3}\int_{r_{\rm min}}^{r_{\rm
max}}\xi(r)r^2dr\ ,
\end{equation}
where we choose $r_{\rm min}=5\ h^{-1}$ Mpc to minimize nonlinear
effects and $r_{\rm max}=20\
h^{-1}$ Mpc to maximize signal-to-noise ratio; within this range of
scales, the model predicted and measured correlation functions are
well approximated by a single power-law.
If we assume $\xi(r)=(r/r_0)^{-\gamma}$, equation (\ref{xi_20})
reduces to
\begin{equation}
\xi_{20}=\frac{3r_0^{\gamma}}{(3-\gamma)r_{\rm max}^3}(r_{\rm
max}^{3-\gamma}-r_{\rm min}^{3-\gamma})\ .
\end{equation}
Because the underlying dark matter correlation function within
this scale range has a power-law index close to $2.0$, we adopt
values from the fixed $\gamma=2.0$ fitting results in Table
\ref{table:fits} instead of the variable power-law index fitting
results. Hence we have $\xi_{20}=1.230\pm 0.353$ for the $2.9\le
z\le 3.5$ bin and $\xi_{20}=2.406\pm 0.586$ for the $z\ge 3.5$
bin, here using the results from all fields.

Our adopted values of $\Phi(z)$ are taken from the
Maximum-Likelihood fitted quasar luminosity function (LF) with
variable power law index given by Richards \etal\ (2006),
integrated from the faintest $i$-band magnitude $i=20.2$.  That
paper uses a slightly different cosmology from our own; we correct
by the ratio of comoving volume elements.
Fig.~20 of Richards \etal\ (2006) shows that the functional fit
we're using here doesn't perfectly follow the data, giving values
of $\Phi(z)$ as much as a factor of 1.5 off from the actual value;
in particular, the variable power law fit function in Richards
\etal\ (2006) appears to underestimate the value of $\Phi(z)$ at
$z<4.5$ but overestimate the value at $z>4.5$ a little bit.  This
will probably cause slight underestimation and overestimation of
$t_{\rm Q}$ (Eq.~\ref{eq:Phi_z}) for the
lower and higher redshift bins respectively, but the effect is
tiny compared with other uncertainties.
Table~\ref{table:LF} lists the values of $\Phi(z)$ we have
calculated, along with other quantities. The limiting absolute
$i$-band magnitude at each redshift is calculated using the same
cosmology and K-correction as in Richards \etal\ (2006),
normalized to $z=2$. One subtlety is that quasars at $z\le 3.0$
are close to the color cut at which the magnitude limit of the
quasar sample changes between $i=19.1$ and 20.2 (see Fig. 17 of
Richards \etal\ 2006). To account for this effect, we use 3 times
the density down to $i=19.1$ for the redshift grid point at
$z=2.9$ and 4 times the density down to $i=19.1$ for the redshift
grid point at $z=3.0$; the grid points with $z \ge 3.1$ use the
integrated luminosity function to $i = 20.2$ (see fig. 17 of
Richards \etal\ 2006).   In practice, our results are insensitive
to these details.

\begin{figure}
  \centering
  \includegraphics[width=0.5\textwidth]{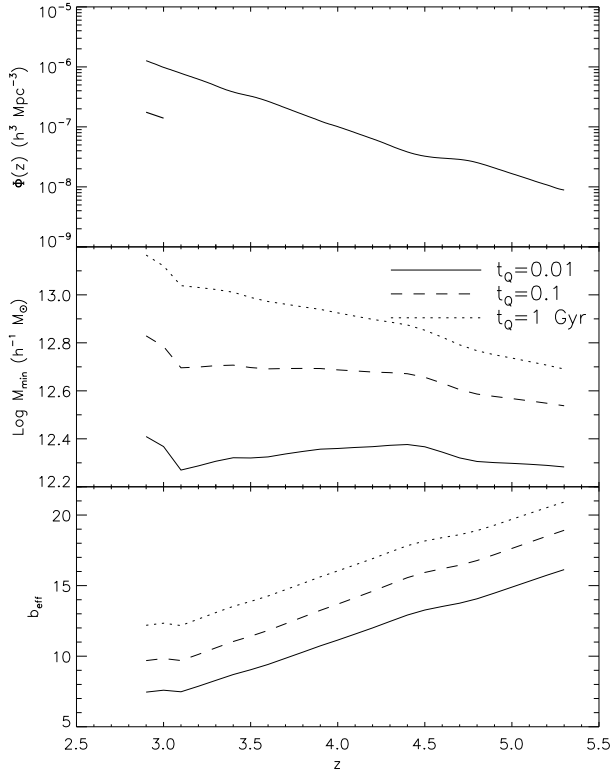}
  \caption{The top panel shows the integrated quasar luminosity
  function (LF) down to the magnitude cut $i=20.2$, computed using the variable power-law fit function
  in Richards \etal\ (2006). The lower line segment shows the integrated LF down to $i=19.1$.
   The bottom two panels show the computed minimum halo masses and effective bias factors as
   functions of redshift, for the three trial values of $t_{\rm Q}=0.01,\ 0.1$ and 1 Gyr.
   We have used the empirical values of $\Phi$ at the grid points $z=2.9$, and $3.0$ (i.e., three and four times the
values down to $i=19.1$, respectively), which causes the jump in
$M_{\rm min}$ and $b_{\rm eff}$ at these two redshift grid points,
i.e., we are targeting more luminous quasars at $z=2.9,\ 3.0$.
   The slight kink around $z=4.5$ in all three panels is due to
   the K-correction (see figure 17 of Richards \etal\ 2006).  }
  \label{fig:tQ_exam}
\end{figure}

To illustrate the relationship between $t_{\rm Q}$, $b_{\rm eff}$,
and $M_{\rm min}$, we choose fixed values of $t_{\rm Q}=0.01,\
0.1,\ 1$ Gyr at each redshift and obtain the corresponding $M_{\rm
min}$ and $b_{\rm eff}$ at $z=3.0,3.5$, and 4.0, listed in
Table~\ref{table:example}. Fig.~\ref{fig:tQ_exam} shows the
evolution of the integrated quasar number density $\Phi(z)$,
$M_{\rm min}(z)$ and $b_{\rm eff}(z)$ for the three trial values
of $t_{\rm Q}$. At each redshift we obtain the model predicted
correlation function $\xi_{\rm model}(r,z)$, which is then
averaged over our sample redshift range weighted by the observed
quasar distribution (not corrected for the selection function)
following equation (\ref{xi_average}).

\begin{deluxetable}{lcccc}
\tablecolumns{5} \tablewidth{0pc} \tablecaption{Trial values of
$t_{\rm Q}$ at redshift $z=3.0, \ 3.5,\ 4.0$ and the corresponding
$M_{\rm min}$ and $b_{\rm eff}$, assuming the fiducial
$\Lambda$CDM cosmology. \label{table:example}} \tablehead{$z$ &
$\Phi$ ($h^3$ Mpc$^{-3}$) & $t_{\rm Q}$ (Gyr) &$M_{\rm min}$
($h^{-1}M_\odot$) & $b_{\rm eff}$} \startdata
3.0 &$5.591\times 10^{-7}$ & 0.01 & $ 2.33 \times 10^{12}$  & 7.6 \\
    &                      & 0.1  & $ 6.10 \times 10^{12}$  & 9.8 \\
    &                      & 1    & $ 1.32 \times 10^{13}$  & 12.3 \\
3.5 &$3.251\times 10^{-7}$ & 0.01 & $ 2.09 \times 10^{12}$  & 9.0 \\
    &                      & 0.1  & $ 4.98 \times 10^{12}$  & 11.4 \\
    &                      & 1    & $ 9.76 \times 10^{12}$  & 13.9 \\
4.0 &$1.009\times 10^{-7}$ & 0.01 & $ 2.29 \times 10^{12}$  & 11.1 \\
    &                      & 0.1  & $ 4.87 \times 10^{12}$  & 13.7 \\
    &                      & 1    & $ 8.41 \times 10^{12}$  & 16.0 \\
\enddata
\end{deluxetable}

\begin{deluxetable*}{lccccccc}
\tablecolumns{8} \tablewidth{\textwidth} \tablecaption{Quasar
space density, $M_{\rm min}$ and $b_{\rm eff}$ at each redshift
grid \label{table:LF}} \tablehead{$z$ &$M_{i, {\rm limit}}$
&$\Phi^{\prime}(M_i<M_{i, {\rm limit}})$ &$\Phi$ & $n_{\rm QSO}$ &
$D(z)$ & $M_{\rm min}$ & $b_{\rm eff}$\\
&$(z=2)$&$h^3\,\rm Mpc^{-3}$&$h^3\,\rm Mpc^{-3}$&$h^3\,\rm
  Mpc^{-3}$&&$h^{-1}\,M_\odot$} \startdata
2.9 &   --         & $4.533\times 10^{-7}$ & $5.268\times 10^{-7}$    & $1.820\times 10^{-7}$ &0.3375 & $3.11\times 10^{12}$& 7.8\\
2.9$^*$ &-26.42    & $1.092\times 10^{-6}$ & $1.268\times 10^{-6}$     & -- & -- & -- & --\\
2.9$^{**}$ &-27.52 & $1.511\times 10^{-7}$ & $1.756\times 10^{-7}$  & -- & -- & -- & --\\
3.0 &--      & $4.808\times 10^{-7}$   & $5.592\times 10^{-7}$   & $2.642\times 10^{-7}$ &0.3293 & $2.81\times 10^{12}$& 8.0\\
3.0$^*$ &-26.51  & $8.445\times 10^{-7}$ & $9.821\times 10^{-7}$     & -- &-- & -- & --\\
3.0$^{**}$ &-27.61  & $1.202\times 10^{-7}$  & $1.398\times 10^{-7}$ & -- &-- & -- & --\\
3.1 &-26.59  & $6.722\times 10^{-7}$ & $7.826\times 10^{-7}$ & $2.735\times 10^{-7}$ &0.3214 &$2.26\times 10^{12}$ & 7.9 \\
3.2 &-26.66  & $5.345\times 10^{-7}$ & $6.228\times 10^{-7}$ & $3.102\times 10^{-7}$ &0.3139 &$2.33\times 10^{12}$ & 8.3\\
3.3 &-26.74  & $4.156\times 10^{-7}$ & $4.847\times 10^{-7}$ & $2.369\times 10^{-7}$ &0.3068 &$2.43\times 10^{12}$ & 8.7\\
3.4 &-26.82  & $3.272\times 10^{-7}$ & $3.820\times 10^{-7}$ & $1.551\times 10^{-7}$ &0.3000 &$2.49\times 10^{12}$ & 9.1\\
3.5 &-26.84  & $2.783\times 10^{-7}$ & $3.251\times 10^{-7}$ & $1.254\times 10^{-7}$ &0.2934 &$2.48\times 10^{12}$ & 9.4\\
\\
3.5 &-26.84  & $2.783\times 10^{-7}$ & $3.251\times 10^{-7}$ & $1.254\times 10^{-7}$ &0.2934 &$5.76\times 10^{12}$ & 11.9\\
3.6 &-26.88  & $2.283\times 10^{-7}$ & $2.670\times 10^{-7}$ & $1.406\times 10^{-7}$ &0.2871 &$5.66\times 10^{12}$ & 12.3\\
3.7 &-26.96  & $1.774\times 10^{-7}$ & $2.076\times 10^{-7}$ & $1.462\times 10^{-7}$ &0.2811 &$5.66\times 10^{12}$ & 12.8\\
3.8 &-27.04  & $1.377\times 10^{-7}$ & $1.612\times 10^{-7}$ & $1.453\times 10^{-7}$ &0.2753 &$5.64\times 10^{12}$ & 13.3\\
3.9 &-27.12  & $1.070\times 10^{-7}$ & $1.254\times 10^{-7}$ & $9.720\times 10^{-8}$ &0.2698 &$5.62\times 10^{12}$ & 13.7\\
4.0 &-27.17  & $8.608\times 10^{-8}$ & $1.009\times 10^{-7}$ & $7.656\times 10^{-8}$ &0.2644 &$5.53\times 10^{12}$ & 14.2\\
4.1 &-27.24  & $6.821\times 10^{-8}$ & $8.002\times 10^{-8}$ & $6.413\times 10^{-8}$ &0.2593 &$5.46\times 10^{12}$ & 14.7\\
4.2 &-27.32  & $5.389\times 10^{-8}$ & $6.326\times 10^{-8}$ & $5.147\times 10^{-8}$ &0.2544 &$5.39\times 10^{12}$ & 15.1\\
4.3 &-27.41  & $4.171\times 10^{-8}$ & $4.898\times 10^{-8}$ & $4.322\times 10^{-8}$ &0.2496 &$5.34\times 10^{12}$ & 15.6\\
4.4 &-27.49  & $3.253\times 10^{-8}$ & $3.823\times 10^{-8}$ & $2.950\times 10^{-8}$ &0.2450 &$5.28\times 10^{12}$ & 16.1\\
4.5 &-27.53  & $2.763\times 10^{-8}$ & $3.248\times 10^{-8}$ & $3.040\times 10^{-8}$ &0.2406 &$5.10\times 10^{12}$ & 16.5\\
4.6 &-27.50  & $2.566\times 10^{-8}$ & $3.018\times 10^{-8}$ & $2.590\times 10^{-8}$ &0.2364 &$4.81\times 10^{12}$ & 16.7\\
4.7 &-27.45  & $2.437\times 10^{-8}$ & $2.867\times 10^{-8}$ & $2.435\times 10^{-8}$ &0.2323 &$4.51\times 10^{12}$ & 17.0\\
4.8 &-27.46  & $2.154\times 10^{-8}$ & $2.535\times 10^{-8}$ & $1.846\times 10^{-8}$ &0.2283 &$4.31\times 10^{12}$ & 17.3\\
4.9 &-27.54  & $1.754\times 10^{-8}$ & $2.066\times 10^{-8}$ & $1.492\times 10^{-8}$ &0.2245 &$4.21\times 10^{12}$ & 17.7\\
5.0 &-27.64  & $1.411\times 10^{-8}$ & $1.662\times 10^{-8}$ & $7.542\times 10^{-9}$ &0.2207 &$4.12\times 10^{12}$ & 18.2\\
5.1 &-27.74  & $1.136\times 10^{-8}$ & $1.339\times 10^{-8}$ & $3.177\times 10^{-9}$ &0.2171 &$4.03\times 10^{12}$ & 18.6\\
5.2 &-27.85  & $9.163\times 10^{-9}$ & $1.080\times 10^{-8}$ & $3.853\times 10^{-9}$ &0.2137 &$3.93\times 10^{12}$ & 19.1\\
5.3 &-27.95  & $7.502\times 10^{-9}$ & $8.847\times 10^{-9}$ & $3.895\times 10^{-9}$ &0.2103 &$3.83\times 10^{12}$ & 19.5\\
\enddata
\tablecomments{\footnotesize $M_{i, {\rm limit}}$ is the $i$ band
limiting absolute magnitude, K-corrected to $z=2$. $\Phi^{\prime}$
is the integrated quasar number density over the apparent
magnitude range, in the same cosmology as in Richards \etal\
(2006), converted using $h=0.7$ to units of $h^3$ Mpc$^{-3}$.
$\Phi$ is the corresponding quasar number density in our
cosmology, converted using $h=0.71$ to $h^3$ Mpc$^{-3}$. There are
three entries for each of the $z=2.9$ and $z=3.0$ grids,
corresponding to a magnitude limit of $i=20.2$ (one asterisk),
$i=19.1$ (two asterisks), and using the empirical values we
adopted at these two redshift grids (see text; no asterisks). The
apparent $i$-band limiting magnitude cut is $i=20.2$ for $z\ge
3.1$. $n_{\rm QSO}$ is the observed overall quasar number density
for all fields, in the current cosmology; the difference between
$n_{\rm QSO}$ and $\Phi$ reflects the selection function and
difference between the fitted power-law function and binned
luminosity function. $D(z)$ is the linear growth factor.
Also tabulated are the corresponding minimal halo mass $M_{\rm
min}$ and effective bias factors $b_{\rm eff}$ at each redshift
grid, computed using the fiducial values of $t_{\rm Q}$,
i.e., $t_{\rm Q}=15$ Myr for $2.9\le z\le 3.5$ and $t_{\rm Q}=160$
Myr for $z\ge 3.5$.}
\end{deluxetable*}
%\clearpage

\begin{figure}
  \centering
  \includegraphics[width=0.5\textwidth]{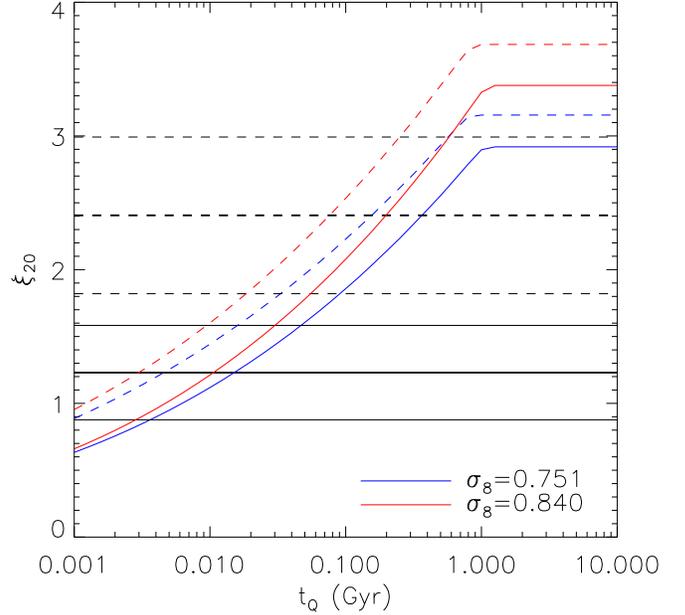}
  \caption{Comparison of the measured and model predicted clustering strength $\xi_{20}$, defined in
  equation (\ref{xi_20}). Solid lines
  correspond to the $2.9\le z\le 3.5$ bin and dashed lines correspond to the $z\ge 3.5$ bin. The
  thick and light horizontal lines show the measured clustering strength and $1-\sigma$ errors.
  The match of the model predicted $\xi_{20}$ (blue lines for the fiducial $\sigma_8=0.751$ and
  red lines for $\sigma_8=0.84$) with the measured $\xi_{20}$ gives the average
  quasar lifetime $t_{\rm Q}$ within that redshift bin. The uncertainty in measured $\xi_{20}$ gives
  a large uncertainty in $t_{\rm Q}$. Quasars in the higher redshift bin have larger
  $t_{\rm Q}$ on average. The fiducial values of $t_{\rm Q}$ inferred from this figure (the
  $\sigma_8=0.751$ case) are: $t_{\rm Q}=15$ Myr for $2.9\le z\le 3.5$ and $t_{\rm Q}=160$ Myr for
  $z\ge 3.5$.}
  \label{fig:xi_20}
\end{figure}

We compare the model predictions and measured values for the $2.9
\le z \le 3.5$ and $z \ge 3.5$ redshift bins respectively.
Fig.~\ref{fig:xi_20} plots the model predicted $\xi_{20}$ as
a function of $t_{\rm Q}$ for the two redshift bins. Above $t_{\rm
Q}\sim 1$ Gyr, the duty cycle saturates at unity, and the
predicted correlation function flattens.  The horizontal lines
show the values and 1$\sigma$ errors of $\xi_{20}$ computed using
our fixed power-law fits, for the two redshift bins respectively.
For the $2.9\le z\le 3.5$ bin, the estimated quasar lifetime is
$t_{\rm Q}\sim 15$ Myr with lower limit 3.6 Myr and upper limit 47
Myr for the 1-$\sigma$ error of the measured $\xi_{20}$. For the
$z\ge 3.5$ redshift bin, the estimated quasar lifetime is $t_{\rm
Q}\sim 160$ Myr with lower limit $\sim 30$ Myr and upper limit
$\sim 600$ Myr for the 1-$\sigma$ error of the measured
$\xi_{20}$.  To phrase this in terms of the duty cycle, we take
the average halo lifetime to be $1$ Gyr at these redshifts (see
Fig. \ref{fig:nm}). Therefore the duty cycle is $0.004\sim 0.05$
for the lower redshift bin and $0.03\sim 0.6$ for the higher
redshift bin.

In the model we are using, $t_{\rm Q}$ is very sensitive to the
clustering strength, as shown in Fig.~\ref{fig:xi_20}. A small
change in the measured quasar correlation function will result in
a substantial change in $t_{\rm Q}$. Using different fitting
results for the measured $\xi_{20}$ (e.g., those for good fields
only) will certainly change the value of $t_{\rm Q}$. However, the
formal 1$-\sigma$ errors of $t_{\rm Q}$ are large enough to
encompass these changes.
The model is also sensitive to the adopted value of $\sigma_8$,
whose consensus value has changed significantly since the release
of the WMAP3 data (Spergel \etal\ 2006). By increasing $\sigma_8$
we can increase the model predicted $\xi_{20}$ given the same
$t_{\rm Q}$\footnote{The $\xi_{20}$ result is insensitive to other
cosmological parameters such as $\Omega_M$.}. The results for the
WMAP first year value $\sigma_8=0.84$ (Spergel \etal\ 2003) are
also plotted in Fig.~\ref{fig:xi_20} as red lines. In this case
the $t_{\rm Q}$ values are slightly lower for the two redshift
bins, but are still within the 1-$\sigma$ errors of the fiducial
$\sigma_8$ case. Combining these effects, we conclude that this
approach can only constrain the quasar lifetime within a very
broad range of $10^6-10^8$ yr, which is, of course, consistent
with many other approaches (e.g., Martini 2004 and references
therein). On the other hand, our results do show, on average, a
larger $t_{\rm Q}$ and duty cycle for the higher redshift bin.

There are other assumptions in our model that we
should consider. In particular, there is the possibility that
quasars cluster more than their dark matter halos due to physical
effects that modulate the formation of quasars on very large
scales. For example, the process of reionization
may show large spatial modulation, which might affect the number
density of young galaxies and quasars on large scales (e.g.,
Babich \& Loeb 2006).
We have also assumed that each halo hosts only one luminous
quasar. However, Hennawi \etal\ (2006a) show that quasars (at
lower redshift) are very strongly clustered on small scales, with
some close binaries clearly in a single halo. Searches for
multiple quasars at higher redshift have also been successful
(Hennawi \etal, in preparation), suggesting that at high redshift
as well, a single halo can host more than one quasar.

Table \ref{table:LF} uses the fiducial values of $t_{\rm Q}$ we
derived for the $\sigma=0.751$ case to estimate the minimal halo
mass and bias factors of high redshift quasars, but the values of
$M_{\rm min}$ and $b_{\rm eff}$ depend only weakly on $t_{\rm Q}$,
as one can see from Table \ref{table:example}.
The values of $M_{\rm min}$ and
$b_{\rm eff}$ are tabulated in Table \ref{table:LF}, for each of
the redshift bins. Note that the change of $M_{\rm min}$ within
each redshift bin may not be real because we have assumed constant
$t_{\rm Q}$ throughout the redshift bin. On the other hand, the
host halos for the higher redshift bin have, on average, a larger
minimal halo mass of $\sim 4-6\times 10^{12}\ M_\odot$ than that
for the lower redshift bin of $\sim 2-3\times 10^{12}\ M_\odot$.
This is expected, because quasars in the higher redshift bin have
higher mean luminosity and hence should reside in more massive
halos. From Table \ref{table:LF} it is clear that high redshift
quasars are strongly biased objects, and the effective bias factor
increases with redshift.

\section{Summary and Conclusions}\label{sec:con}
We have used $\sim 4000$ high redshift SDSS quasars to measure the
quasar correlation function at $z\ge 2.9$. The clustering of these
high redshift quasars is stronger than that of their low redshift
counterparts. Over the range of $4 < r_p < 150\ h^{-1}$ Mpc, the
real-space correlation function is fitted by a power-law form
$\xi(r)=(r/r_0)^{-\gamma}$ with $r_0\sim 15\ h^{-1}$ Mpc and
$\gamma\sim 2$. When we divide the clustering ample into two broad
redshift bins, $2.9\le z\le 3.5$ and $z\ge 3.5$, we find that the
quasars in the higher redshift bin show substantially stronger
clustering  properties, with a comoving correlation length
$r_0=24.3\pm 2.4\ h^{-1}$ Mpc assuming a fixed power-law index
$\gamma=2.0$. The lower redshift bin has a comoving correlation
length $r_0=16.9\pm 1.7\ h^{-1}$ Mpc, assuming the same power-law
index.

We followed Martini \& Weinberg (2001) to relate this strong
clustering signal to the quasar luminosity function (Richards
\etal\ 2006), the quasar lifetime and duty cycle, and the mass
function of massive halos. We find the minimum mass $M_{\rm min}$
of halos in which luminous quasars in our sample reside, as well
as the clustering bias factor for these halos. High redshift
quasars are highly biased objects with respect to the underlying
matter, while the minimal halo mass shows no strong evolution with
redshift for our flux-limited sample. Quasars with $2.9\le z\le
3.5$ reside in halos with typical mass $\sim 2-3\times 10^{12}\
h^{-1}\ M_\odot$; quasars with $z\ge 3.5$ reside in halos with
typical mass $\sim 4-6\times 10^{12}\ h^{-1}\ M_\odot$. The slight
difference of $M_{\rm min}$ in the two redshift bins is expected
because quasars in the higher redshift bin have mean luminosity
that is approximately two times that of quasars in the lower
redshift bin, and should reside in more massive halos. We further
estimated the quasar lifetime $t_{\rm Q}$. We get a $t_{\rm Q}$
value of $4\sim 50$ Myr for the $2.9\le z\le 3.5$ bin and $30\sim
600$ Myr for the $z\ge 3.5$ bin; which is broadly consistent with
the quasar lifetime of $10^6-10^8$ yr estimated from other methods
(e.g., Martini 2004 and references therein).  This corresponds to
a duty cycle of $0.004\sim 0.05$ for the lower redshift bin and
$0.03\sim 0.6$ for the higher redshift bin, where we take the
average halo lifetime to be $1$ Gyr. In general we find the
average lifetime is higher for the higher redshift bin, which
could either be due to the redshift evolution or an effect of the
luminosity dependence of $t_{\rm Q}$. However, we emphasize that
our approach is subject to a variety of uncertainties, including
errors in the clustering measurements themselves, uncertainties in
$\sigma_8$ and the halo mass function, and the validity of the
assumptions we have adopted.

It is interesting to note that recent {\it Chandra} and {\it
XMM-Newton} studies on the clustering of X-ray selected AGN have
revealed a larger correlation length than optical AGN.  In
particular, hard X-ray AGN have a correlation length $r_0\sim 15\
h^{-1}$ Mpc at $z\lesssim 2$ (e.g., Basilakos \etal\ 2004; Gilli
\etal\ 2005; Puccetti \etal\ 2006; Plionis 2006). Given the fact
that X-ray selected AGN have considerably lower mean bolometric
luminosity than do optically-selected AGN (e.g., Mushotzky 2004),
this implies, once again, that the instantaneous luminosity is not
a reliable indicator of the host halo mass at the low luminosity
end (e.g., Hopkins \etal\ 2005).  Shen \etal\ (2007) have
suggested an evolutionary model of AGN accretion in which an AGN
evolves from being dominant in the optical to dominant in X-rays
when the accretion rate drops. Hence those strongly clustering
hard X-ray AGN were probably once very luminous quasars in the
past with high peak luminosities. When they dim and turn into hard
X-ray sources, their spatial clustering strength remains. However,
the current X-ray AGN sample is still very limited compared with
optically selected samples, hence the uncertainty in the X-ray AGN
correlation length is large.

The work described in this paper can be extended in a variety of ways.  Our sample
cannot explore clustering below $\sim 1\ h^{-1}$ Mpc because of fiber
collisions; we are extending the methods of Hennawi \etal\ (2006a) to
find close pairs of high-redshift quasars, to determine whether the
excess clustering found at moderate redshift extends to $z > 3$.
Extending the clustering analysis to lower luminosities will be
important, given theoretical predictions of a strong luminosity
dependence to the clustering signal at high redshifts (Hopkins \etal\
2006).  The repeat scans of the Southern Equatorial Stripe in SDSS
(Adelman-McCarthy \etal\ 2007) will allow us to extend the luminosity
range of our sample, and redshifts of the fainter quasars are already
being obtained (Jiang \etal\ 2006).  The massive halos that we
predict host the luminous quasars must also contain a substantial
number of ordinary galaxies, and we plan deep imaging surveys of
high-redshift quasar fields to measure the quasar-galaxy crosss-correlation
function (see Stiavelli \etal\ 2005; Ajiki \etal\ 2006).  Finally,
more work is needed on simulations of quasar clustering.  Our quasar
lifetime/duty cycle calculation is frustratingly imprecise, and
further explorations of the behavior of highly biased rare halos at
high redshifts may yield ways to constrain duty cycles more directly
from the data, and understand the uncertainties of the technique in
more detail.

Finally, we need to make more detailed comparisons of high-redshift
quasar clustering with that of luminous galaxies at the same
redshift.  The duty cycle of quasars at these redshifts is a few
percent at most, thus there is a population of galaxies with quiescent
central black holes that is just as strongly clustered.  The
correlation length of Lyman-break galaxies at these redshifts is $\sim
5\,h^{-1}$Mpc (Adelberger \etal\ 2005a), but the clustering strength
appears to increase (albeit at $z \sim 2$) with increasing observed K-band
luminosity (Adelberger \etal\ 2005b; Allen \etal\ 2005) and/or color (Quadri \etal\
2006).  The duty cycle we have calculated should agree with the ratio
of number densities of luminous quasars, and that of the parent host
galaxy population.  The challenge will be to identify this parent
population unambigously.

\acknowledgements We thank Paul Bode and J. P. Ostriker for useful
discussions and for providing us their numerical simulation
results used in Appendix \ref{app:martini_weinberg}, and Jim Gray
for his work on the CAS. YS and MAS acknowledge the support of NSF
grant AST-0307409.

Funding for the SDSS and SDSS-II has been provided by the Alfred
P. Sloan Foundation, the Participating Institutions, the National
Science Foundation, the U.S. Department of Energy, the National
Aeronautics and Space Administration, the Japanese Monbukagakusho, the
Max Planck Society, and the Higher Education Funding Council for
England. The SDSS Web Site is http://www.sdss.org/.

The SDSS is managed by the Astrophysical Research Consortium for the
Participating Institutions. The Participating Institutions are the
American Museum of Natural History, Astrophysical Institute Potsdam,
University of Basel, University of Cambridge, Case Western Reserve
University, University of Chicago, Drexel University, Fermilab, the
Institute for Advanced Study, the Japan Participation Group, Johns
Hopkins University, the Joint Institute for Nuclear Astrophysics, the
Kavli Institute for Particle Astrophysics and Cosmology, the Korean
Scientist Group, the Chinese Academy of Sciences (LAMOST), Los Alamos
National Laboratory, the Max-Planck-Institute for Astronomy (MPIA),
the Max-Planck-Institute for Astrophysics (MPA), New Mexico State
University, Ohio State University, University of Pittsburgh,
University of Portsmouth, Princeton University, the United States
Naval Observatory, and the University of Washington.

Facilities: \facility{Sloan}

\clearpage

\appendix
\section{A. Quasar Redshift Determination}
\label{app:redshift}

\subsection{A.1 Broad Emission Line Shifts}
High redshift quasars ($z\ge2.9$) have only a few strong emission
lines that fall within the SDSS spectral coverage (3800-9200 \AA):
Ly{\sevenrm $\alpha$} (1216 \AA), \SiIV/\OIV\ (1397 \AA), \CIV\
(1549 \AA) and \CIII\ (1909 \AA). The Ly$\alpha$ emission line is
heavily affected by the Lyman $\alpha$ forest, and is blended with
N{\sevenrm V} 1240 {\AA}. In addition, high-ionization broad
emission lines such as \CIV\ are blueshifted by several hundred km
s$^{-1}$ from the redshift determined from narrow forbidden lines
like [OIII]5007 \AA\ (e.g., Gaskell 1982; Tytler \& Fan 1992;
Richards \etal\ 2002a).  We could simply correct the redshift
derived from each observed line for the (known) mean offset of
that line from systemic (e.g., Vanden Berk \etal\ 2001; Richards
\etal\ 2002a). We can do better than this, however, by examining
the relationships between the shifts of different lines.

To understand these relationships, we use a sample of quasars
drawn from the SDSS DR3 quasar catalog (Schneider \etal\ 2005)
with $1.8 \le z \le 2.2$; for these objects, the lines \SiIV,
\CIV, \CIII\ and \MgII2800 \AA\ all fall in the SDSS spectral
coverage.  The \MgII\ line has a small and known offset from the
systemic redshift (Richards \etal\ 2002a), thus tying our results
to \MgII\ allows us to determine the systemic redshift for each
object. We exclude from the sample those objects which show
evidence for a broad absorption line, determined using the
``balnicity'' index (BI) of Weymann \etal\ (1991) and using the
Vanden Berk \etal\ (2001) quasar composite spectrum to define the
continuum level.

\begin{figure}
  \centering
    \includegraphics[scale=0.6]{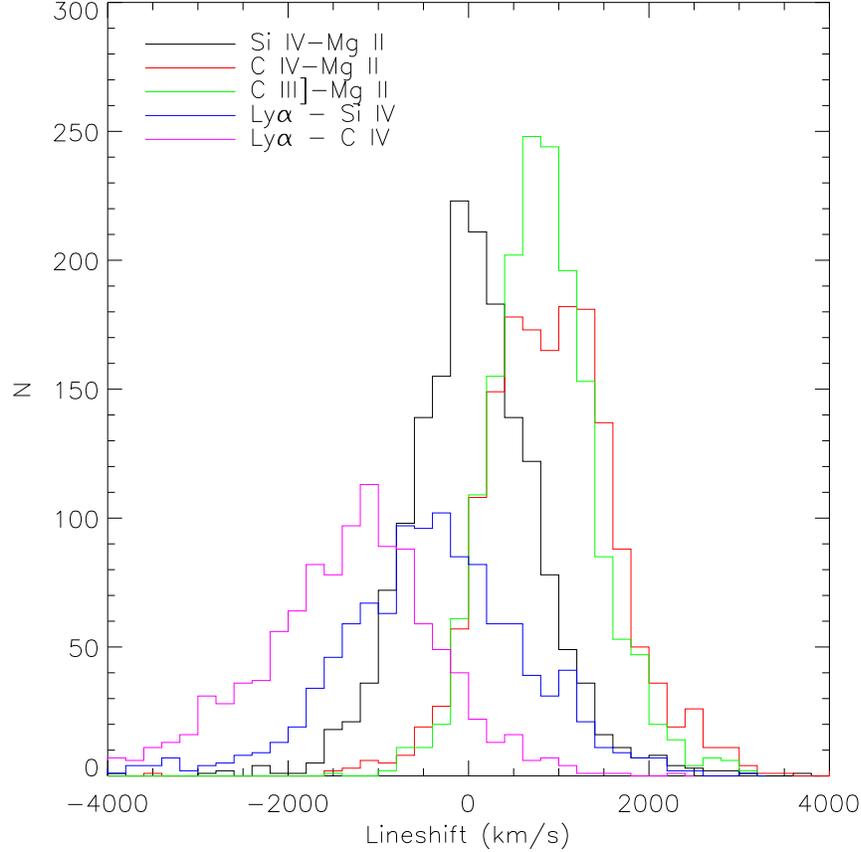}
    \caption{Distributions of relative shifts of the modes of various emission lines,
     as measured for 1652 high S/N, non-BAL quasars with
      redshifts between 1.8 and 2.2. The mean values and 1$-\sigma$ deviations of these
      line shifts are listed in Table \ref{table:lineshift}. }
    \label{fig:lineshift_histogram}
\end{figure}

We fit a log-normal to each of the four lines (with a second
log-normal added for the neighboring lines He{\sevenrm II}1640
\AA\ and Al{\sevenrm III}1857 \AA), together with the local
continuum. The centroid for each line is determined following
Hennawi \etal\ (2006b): we calculate the mode of the pixels within
$\pm 1.5\sigma$ of the fitted Gaussian line center using
$3\times{\rm median}-2\times{\rm mean}$.
We include in the mode calculation those pixels with flux:
\begin{equation}
f_{\lambda}>\frac{0.6A_i}{\sqrt{2\pi
\sigma_i}}+C_\lambda+\sum_{j\neq
i}\frac{A_j}{\sqrt{2\pi\sigma_j}}e^{-(\log_{10}\lambda-\log_{10}{\lambda_j})^2/2\sigma_j^2}\
,
\end{equation}
where $A_i$, $\log_{10}\lambda_i$ and $\sigma_i$ are the
amplitude, central wavelength, and dispersion of the best fit
log-normal to the $i$th emission line and $C_\lambda$ is the
linear continuum.  Lines with a
  signal-to-noise ratio (S/N) less than 6 per pixel, or with log-normal fits
  with $\chi^2 > 5$ are rejected from further consideration.  This
  gives us a sample of 1652 quasars with robust line measurements.
  Fig.~\ref{fig:lineshift_histogram} shows the distribution of
  shifts between various lines.  The means and standard deviations of
  these distributions are given in Table~\ref{table:lineshift}.  The
  contribution from the line fitting error is
negligible compared to the ``intrinsic'' dispersion of velocity
shifts.

\begin{figure}
  \centering
    \includegraphics[width=0.45\textwidth]{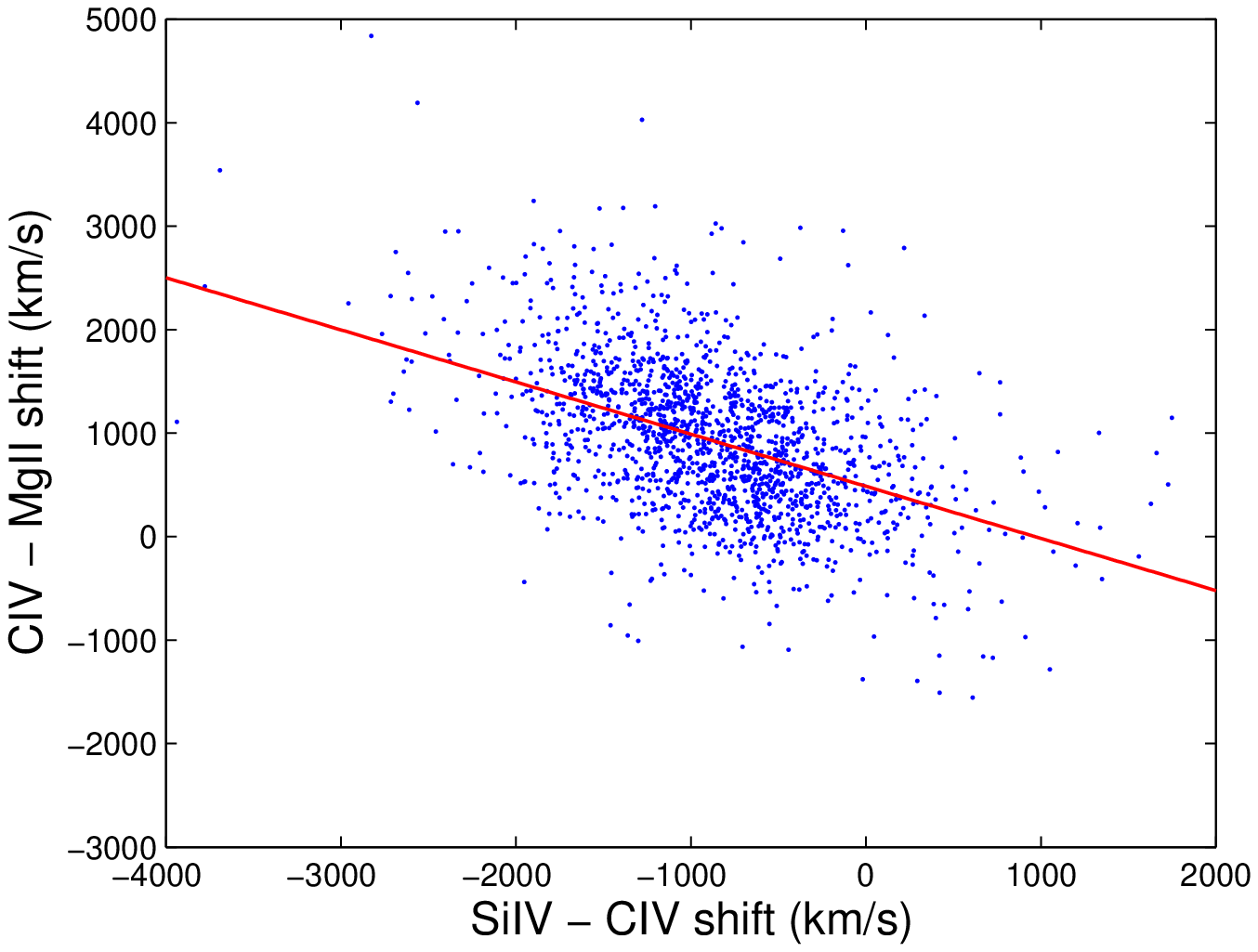}
    \includegraphics[width=0.45\textwidth]{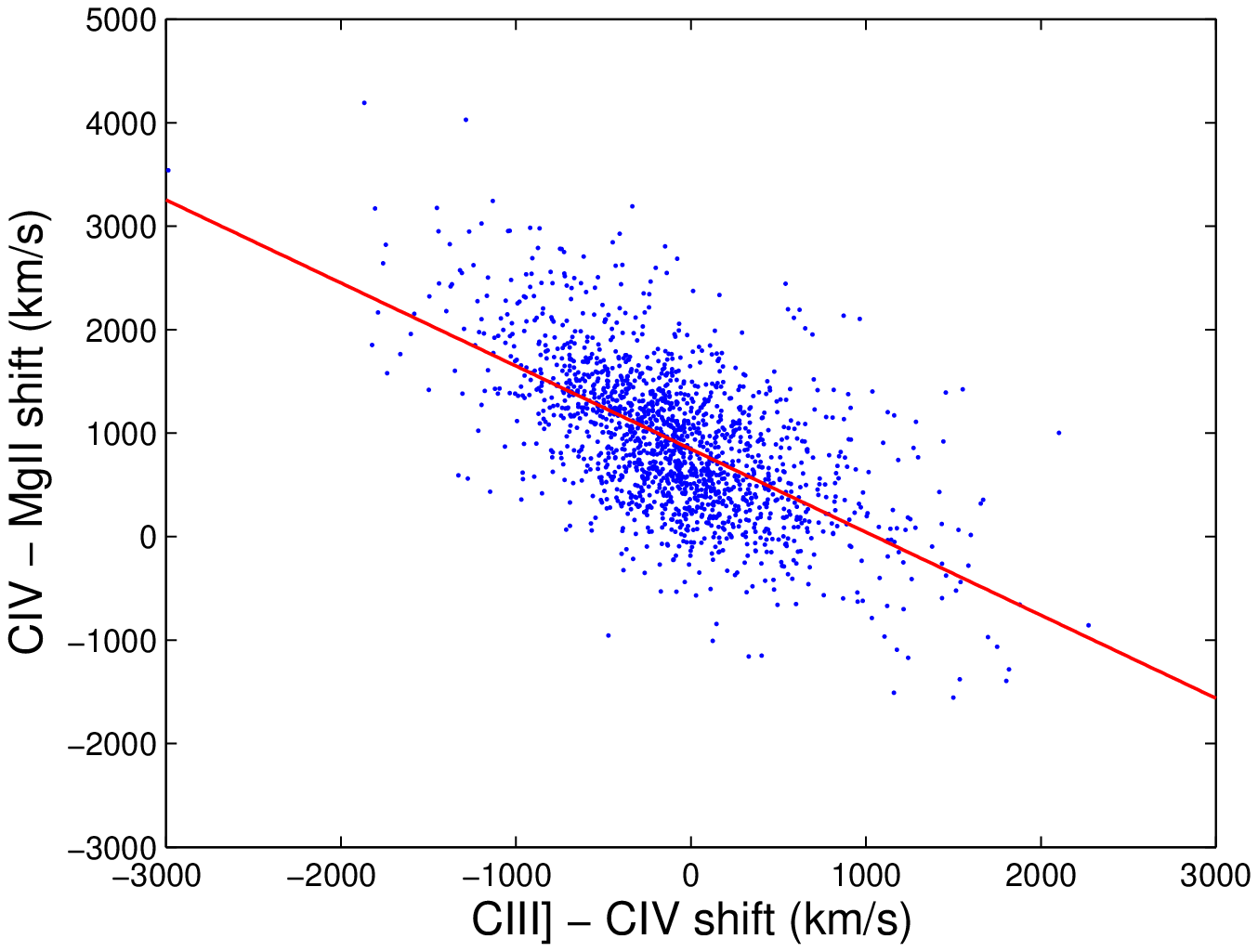}
    \includegraphics[width=0.45\textwidth]{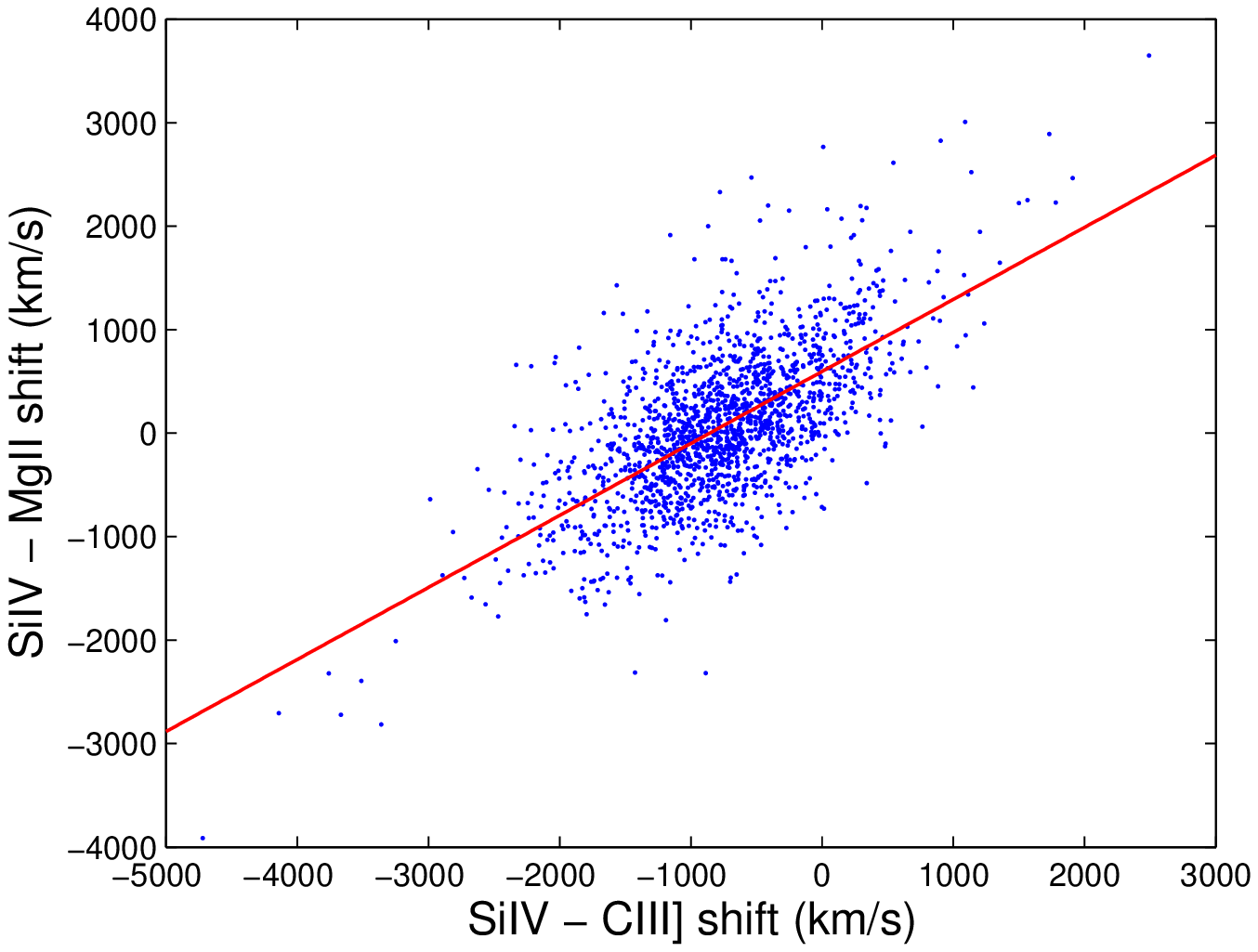}
    \caption{Correlations between various emission line shifts. Blue dots are data points and
    red lines are fitted linear functions. These correlations
    are used in our redshift estimation. The fitted linear parameters and 1$-\sigma$ deviations
    are listed in Table \ref{table:lineshift}.}
    \label{fig:lineshift_correlation}
\end{figure}

These line shifts are correlated with each other, as
Fig.~\ref{fig:lineshift_correlation} shows.  In each panel, we
show the best-fit line to the correlations, giving each point
equal weight. Given these correlations, we can use the shifts
between the lines we observe at high redshift to determine the
offset to \MgII, and thus to the systemic redshift.

There are also correlations between the lineshifts and quantities
such as the quasar luminosity, color, line width, and equivalent
width. However, these correlations show large scatter, and are
therefore not as good for determining the true redshifts of the
quasars.

\begin{deluxetable*}{lcccccc}
%\rotate
\tablecolumns{7} \tablewidth{\textwidth} \tablecaption{Emission
line shifts} \tablehead{  & Ly{\sevenrm $\alpha$} - \SiIV\ &
Ly{\sevenrm $\alpha$}-\CIV & \SiIV\ - \MgII\ &
  \CIV\ - \MgII  & \CIII\ - \MgII & \MgII\ - [OIII]} \startdata
mean vel shift (km s$^{-1}$) & -463   & -1478 &  61 & 921 & 827 & -97\\
$\sigma$  (km s$^{-1}$)      & 1178   & 1217  & 744 & 746 & 604 & 269\\
\hline\\
$y=ax+b$  & $a$  & & $b$ (km s$^{-1}$)  & & $\sigma$ (km s$^{-1}$) \\
\CIV\ - \MgII\ vs. \SiIV\ - \CIV\ & -0.5035 & & 486.7 & & 660\\
\CIV\ - \MgII\ vs. \CIII\ - \CIV\ & -0.8024 & & 845.8 & & 594\\
\SiIV\ - \MgII\ vs. \SiIV\ - \CIII\ & 0.6958 & & 596.5 & & 569\\
\enddata
\tablecomments{\footnotesize The \MgII\ - [OIII] (i.e., systemic)
  lineshift and 1$\sigma$ error are taken from Richards \etal\
(2002).  Positive values indicate a blueshift. \\
The dispersion of the shift between \CIV\ and \MgII\ is somewhat
larger than the value of 511 km s$^{-1}$ quoted by Richards \etal\
(2002), but is consistent with their recent result using a much
larger sample from SDSS DR4 ($\sim$ 770 km s$^{-1}$).
  }
\label{table:lineshift}
\end{deluxetable*}

\begin{figure}
  \centering
    \includegraphics[width=0.45\textwidth]{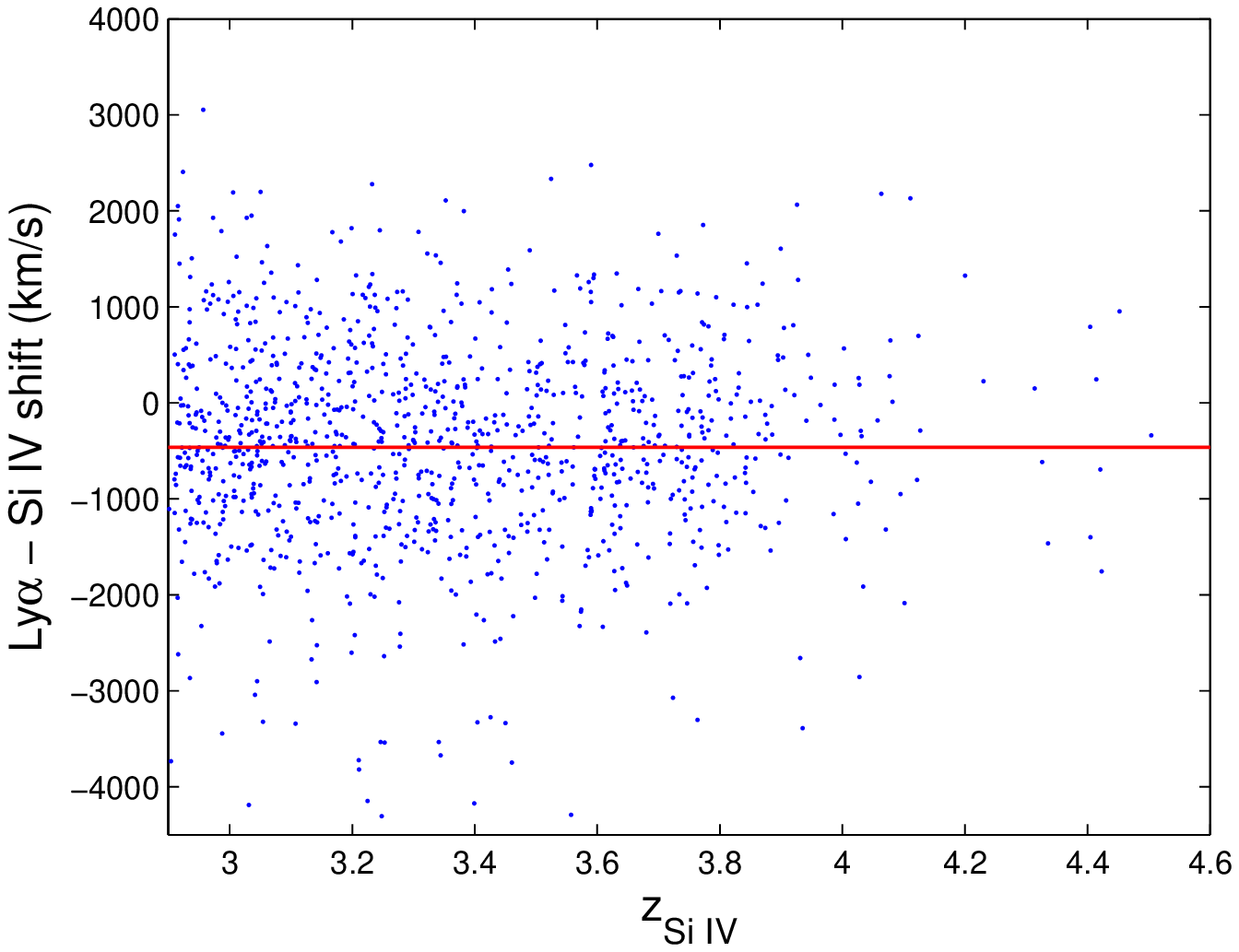}
    \includegraphics[width=0.45\textwidth]{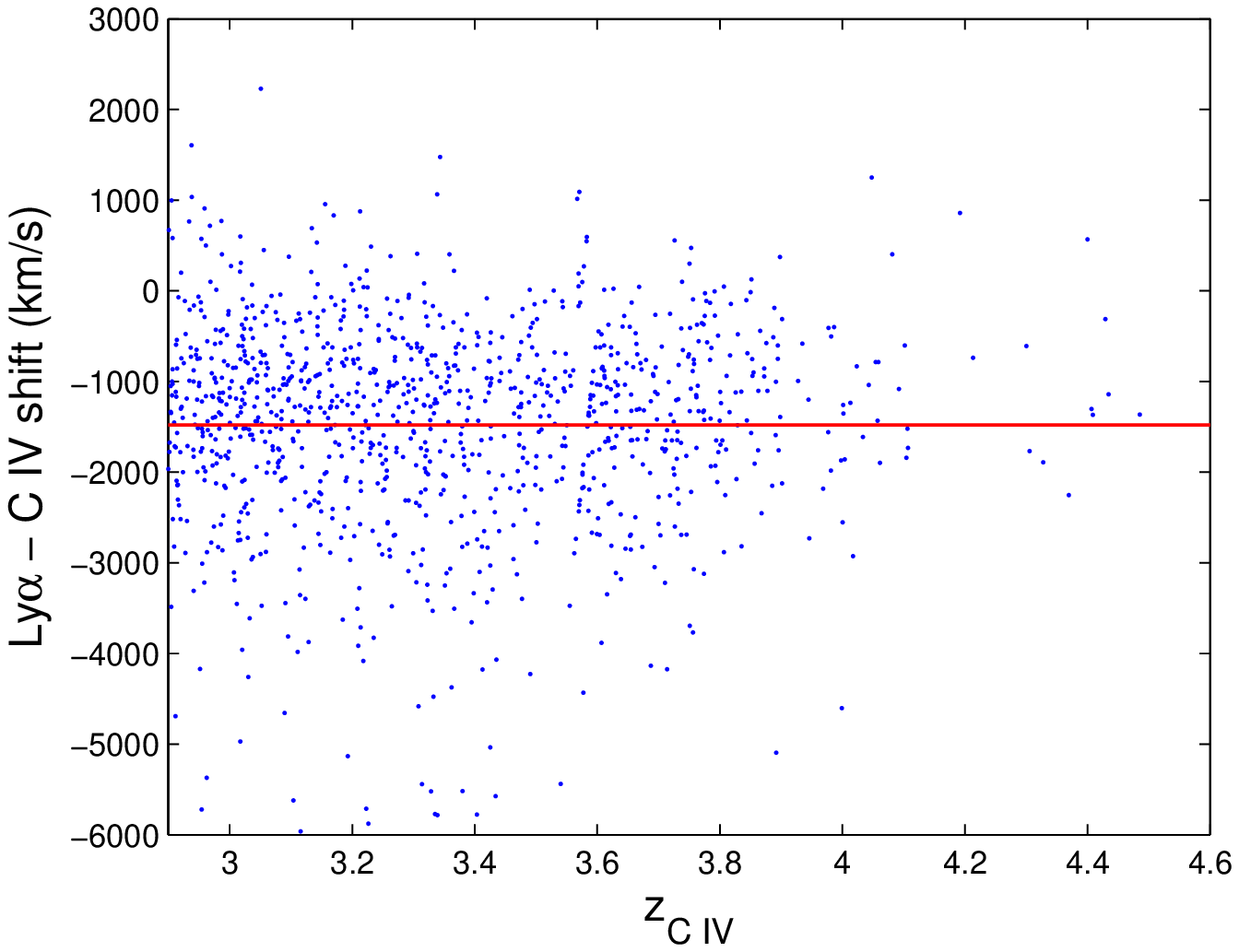}
    \caption{Relative shifts of Ly $\alpha$ versus \SiIV\ and \CIV\ emission
    lines as a function of redshift. Red lines indicate the mean value
    of line shifts. The mean values of line shifts and 1$-\sigma$
    deviations are listed in Table \ref{table:lineshift}.}
    \label{fig:lyalpha_lineshift}
\end{figure}

\subsection{A.2 Ly {\sevenrm $\alpha$} $-$ \SiIV, Ly {\sevenrm $\alpha$} $-$ \CIV\ Line Shifts}

The \CIV\ line lies beyond the SDSS spectra for $z > 4.9$. In
addition, some quasars have weak metal emission lines, which are
of too low S/N to allow us to measure a redshift from them.  In
these cases, we will measure the redshift from the Ly$\alpha$
line.  In order to understand the biases that this gives, we
selected a sample of 1114 non-BAL quasars with $2.9 < z < 4.8$
with high S/N \SiIV\ and \CIV\ lines.  The center of the
Ly$\alpha$ line was taken to be the wavelength of maximum flux. To
reduce the effects of fluctuations and strong skylines, we mask
out 5-$\sigma$ outliers from the 20-pixel smoothed spectrum and
the 5577 \AA\ skyline region (about 20 pixels), and smooth the
spectrum by 15 pixels before identifying the peak pixel; all
spectra were examined by eye to confirm that we correctly
identified the peak of Ly$\alpha$.

Fig.~\ref{fig:lyalpha_lineshift} shows the shifts between
Ly$\alpha$ and the \CIV\ and \SiIV\ lines as a function of
redshift.  The mean shift is $\sim 500$ km s$^{-1}$, with a
1$\sigma$ scatter of 1200 km s$^{-1}$ for Ly$\alpha$-\SiIV; and is
$\sim 1500$ km s$^{-1}$, with a 1$\sigma$ scatter of 1200 km
s$^{-1}$ for Ly$\alpha$-\CIV. This systematic offset is caused by
absorption blueward of the Ly$\alpha$ forest; over this redshift
range, the increasing strength of the forest doesn't cause an
appreciable increase in the shift. The Ly{\sevenrm $\alpha$} line
is blended with the N{\sevenrm V} line, therefore whenever we use
Ly{\sevenrm $\alpha$} as the only estimator for redshift, we
examine the spectrum by eye to confirm that we have identified the
correct line.

\subsection{A.3 Determination of Redshifts}

We are now ready to determine unbiased redshifts for our sample of $z
\ge 2.9$ quasars.
Given the first guess of the redshift of each object from
Schneider \etal\ (2005) for those objects included in DR3, and
from the two spectroscopic pipelines (\S~\ref{sec:parent_sample}),
we fit the centroids of the \SiIV, \CIV\ and \CIII\ lines as we
described above.

For objects in which the centroids of all three lines are
well-determined (we require that a line have a mean S/N per pixel $> 4$ and
  reduced $\chi^2 < 10$), we base the redshift on the centroid of
  \CIV.  We measure the shift between \CIV\ and \SiIV, and the shift
  between \CIV\ and \CIII, and determine from each the expected
  \CIV-\MgII\ line shift using the correlations in
  Fig.~\ref{fig:lineshift_correlation} and
  Table~\ref{table:lineshift}.  We average these lineshifts together,
  and add on the small correction from \MgII\ to systemic given by
  Richards \etal\ (2002a); this gives our final \CIV\ to systemic shift and hence the redshift.
  The uncertainty in these shifts gives rise to an uncertainty $\sigma_v
=
  519 {\rm\ km\ s}^{-1}$ or $\sigma_z = (1 + z)\sigma_v/c$.

For quasars with only two high S/N lines, we take \CIV\ whenever
we have it and \SiIV\  when \CIV\ is absent (we avoid using \CIII\
because it is often near the upper wavelength limit, $9200$ \AA,
of the SDSS spectra).  Again, we use the correlations of
Fig.~\ref{fig:lineshift_correlation} to compute the line shift
relative to \MgII\ and therefore the shift relative to the
systemic redshift. The velocity shift (relative to systemic)
errors in this correction are: 713 km s$^{-1}$ if the two lines
are \SiIV\ and \CIV; 629 km s$^{-1}$ if the two lines are \SiIV\
and \CIII, and 652 km s$^{-1}$ if the two lines are \CIV\ and
\CIII. For quasars with only one well-detected line, we use the
average line shift, and use error transfer to determine the errors
in the line shift relative to systemic. These errors are: 791 km
s$^{-1}$ for \SiIV, 793 km s$^{-1}$ for \CIV\ and 661 km s$^{-1}$
for \CIII.
Finally, for those quasars with no well-detected metal lines, we
use Ly$\alpha$ to determine the redshift, using the average line
shift relative to \CIV\ and the corresponding 1-$\sigma$
dispersion to compute the error: adding the uncertainties in the
transformations in quadrature gives an error of 1453 km s$^{-1}$.

Finally, we examine the spectra of the following classes of
objects by eye to check the redshift determinations: (1) the 407
objects with $|z_{\rm i}-z_{\rm sys}|>3\sigma_z$, where $z_{\rm
i}$ is the initial redshift from the DR3 QSO catalog or SDSS
spectroscopic pipeline; $z_{\rm sys}$ is our best estimation of
redshift and $\sigma_z$ is the estimated redshift
error;
(2) the 327 objects for which the redshift was based on
Ly$\alpha$; and (3) serendipitously found ambiguous cases.
Of the $\sim 750$ objects we inspected by eye, our redshift as
determined above was superior to the value from Schneider \etal\
(2005) or the pipelines in 70\% of the quasars; for 15\%, at least
one of the pipeline redshifts was correct and was therefore
adopted, and for the remaining 15\% (many of them are BAL),
neither redshift was correct. In the latter case, we refit the
redshift by hand, and assigned a redshift error $\sigma_z$ between
0.01 and 0.05, depending on how messy the spectrum was.  There
were 29 objects whose redshifts were undetermined, lay below 2.9,
or were simply not quasars. Thus the parent sample, from which we
will construct our clustering subsample, contains 6,109 objects
(including $\sim 200$ duplicates).

Finally, we compared the redshifts in our sample with the
separately compiled DR5 quasar sample of Schneider \etal\ (2006).
The difference in redshifts follows a Gaussian distribution with
zero mean and a dispersion of 0.01, comparable to our estimated
errors.

\section{B. Survey Geometry}
\label{app:geometry}

\begin{figure}
  \centering
    \includegraphics[scale=1]{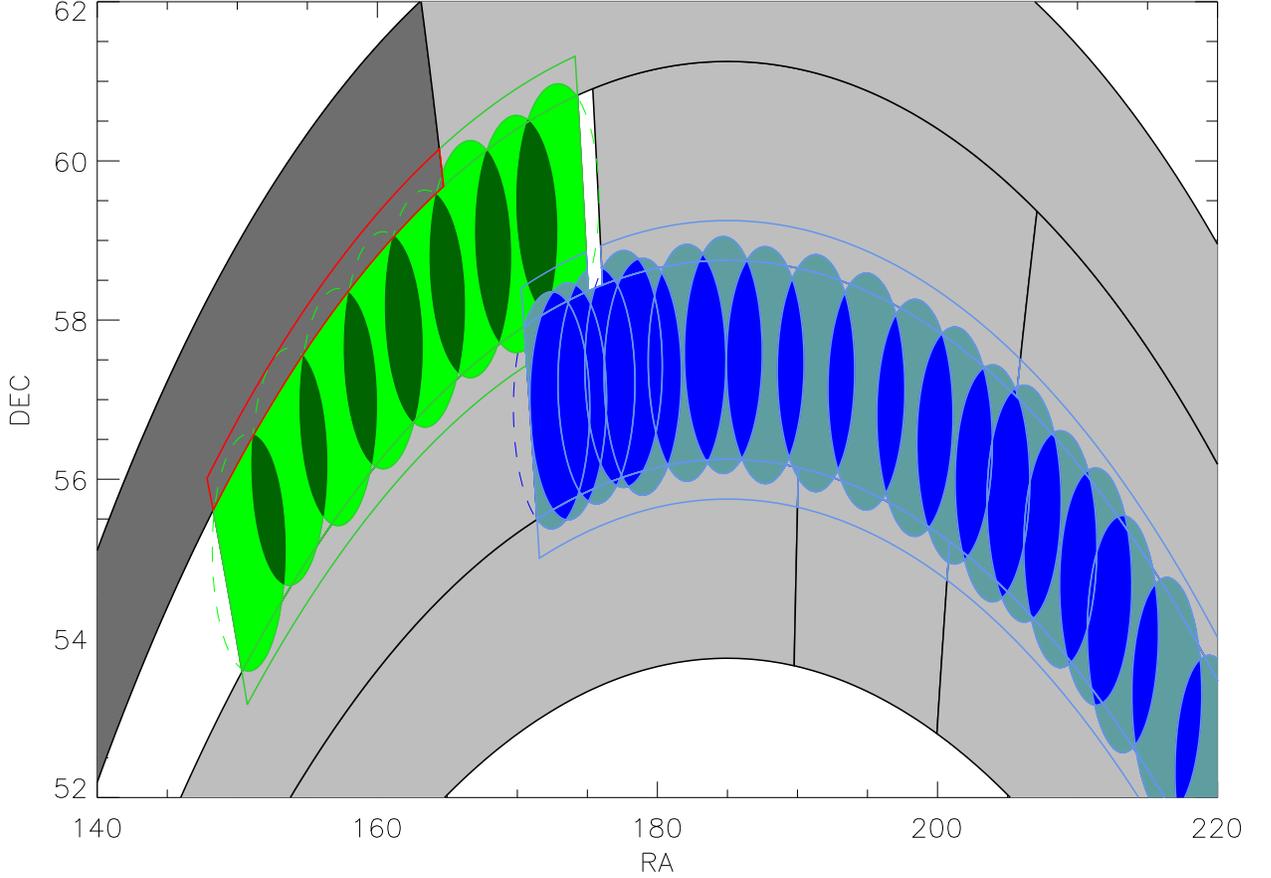}
    \\[-10pt]
    \caption{\small{Portion of the targeting and tiling geometry in SDSS spectroscopy.
    The targeting chunks are denoted by stripes bounded by black lines and each targeting chunk
    is targeted using one target version. Gray stripes are targeting chunks with target version
    no lower than v3\_1\_0 (not necessarily the same version); one dark gray targeting chunk shown here is
    targeted with target version v2\_13\_5. Within targeting chunks we carve out tiling
    rectangles, each of which is targeted with a unique version. A set of tiling rectangles form
    a tiling chunk. Shown here as examples are tiling chunk 38, which has one rectangle (red) targeted
    with version v2\_13\_5 and three rectangles (green) with version v3\_1\_0; tiling chunk 67,
    whose rectangles (blue) are all with target version v3\_1\_0 or later. Within each tiling
    chunk we place tiles ($1^\circ.49$ radius circles, which appear as ellipses because the aspect ratio of the region of
    sky shown is not $1:1$); tiles are trimmed by the boundaries of
    rectangles of that tiling chunk and balkanized (i.e., Hamilton \& Tegmark 2004) into non-overlap
    sectors (which are covered by only one tile) and overlap sectors (which are covered by more than
    one tile). We use light and dark colors to denote the two types of sectors in the above two tiling
    chunks. Note that though balkanized sectors of the same tiling chunk do not intersect with each other,
    they could intersect with sectors of another tiling chunk. In the above case, the
    upper-left corner rectangle in tiling chunk 67 is completely within the middle main rectangle of
    tiling chunk 38. Therefore one should be careful when computing the effective area of sectors.
    In constructing our clean subsample for clustering analysis, we reject those sectors that are
    within tiling rectangles which are targeted with target version lower than v3\_1\_0, i.e., regions
    such as the red rectangle in chunk 38.}}
    \label{fig:geometry}
\end{figure}

SDSS spectroscopic targets are selected from the imaging data, and
thus the spectroscopic footprint is a complicated combination of
the individual runs which make up the imaging data, and the
circular $1.49^\circ$ radius tiles on which spectroscopic targets
are assigned to fibers.  Here we describe how this footprint is
quantified.  It will be useful in the following discussion to
refer to Fig.~\ref{fig:geometry}. Related discussions of the SDSS
survey footprint in the context of galaxy samples may be found in
Appendix A2 of Tegmark \etal\ (2004) and in Blanton \etal\ (2005).

As described in York \etal\ (2000), each imaging run of the SDSS
covers part of a strip; two adjacent strips make a filled {\em
stripe} of width $2.5^\circ$.
Spectroscopic targeting to define a set of tiles is done off
contiguous pieces of stripes termed {\em targeting chunks}; the
SDSS imaging never got so far ahead of the spectroscopy to allow a
targeting chunk to work off more than one stripe at a time.  The
targeting in a given chunk all uses the same version of the target
selection code (an important consideration for us, given the
change in quasar target selection following DR1;
\S~\ref{sec:cluster_sample}). Each targeting chunk is bounded on
the East and West by lines of constant $\mu$ (i.e., the SDSS great
circle coordinate; see Pier \etal\ 2003), and, for stripes in the
Northern Galactic Cap, they are bounded in the North-South
direction by lines of constant $\eta$ (i.e., the SDSS survey
coordinate) if in the Northern stripes. Targeting chunks in the
three stripes in the Southern Galactic Cap have no $\eta$ boundary
applied.
Targeting chunks never overlap, therefore the union of targeting
chunks defines the geometry of the targeting regions as a whole.
Parameters defining the geometry of the targeting chunks can be
found in a table called {\tt Chunk}\footnote{We used the TARGET
(not BEST) version of the {\tt Chunk} table.} in the CAS.

As described by Blanton \etal\ (2003), targets in each chunk are
assigned to tiles, and then to fibers within each plate.   We
first define {\it tiling chunks} (referred to as ``tiling
regions'' by Blanton \etal\ 2003; Blanton \etal\ 2005), each of
which is a set of non-overlapping tiling rectangles bounded by
constant coordinates in different coordinate systems (all three
types of coordinate systems, as well as the mixture of them are
used in describing the tiling rectangles; and there is a flag
indicating the coordinate type in the {\tt TilingBoundary} table
in the CAS). Each of these tiling rectangles lies completely within a
single targeting chunk so that the target selection version is
unique throughout the rectangle.

Although tiling rectangles of the same tiling chunk never overlap,
tiling rectangles from different tiling chunks can overlap; for
example, the upper-left blue rectangle and the middle main green
rectangle in Fig.~\ref{fig:geometry}. On the other hand, a tiling
rectangle never straddles two {\em targeting} chunks, so the
target selection version is the same over the rectangle. A tiling
chunk as a whole can straddle more than one targeting chunk, and
can have tiling rectangles that don't all use the same version
of the target selection pipeline.
A set of spectroscopic tiles of radius $1^\circ.49$ are placed in
each tiling chunk, and fibers assigned to the targeted objects
therein, following the algorithm of Blanton \etal\ (2003). Thus
because the tiles often extend beyond the boundaries of the tiling
chunk (see Fig.~\ref{fig:geometry}), they do not include any
targets beyond the tiling chunks.
The intersection of the tiling rectangles and the circular tiles
defines {\it sectors}: each sector is covered by a unique set of
tiles (see Figure 3 of Blanton \etal\ 2005), and is a
spherical polygon as described by Hamilton \& Tegmark (2004).
The union of all the sectors defines the angular coverage of the
SDSS. We say a sector is a {\it ``non-overlap sector''} if it is
covered by only one tile (the lighter colors in
Fig.~\ref{fig:geometry}) and is an {\it ``overlap sector''} if it
is covered by more than one tile (indicated with darker colors in
the figure).

The tiling chunk geometry information is taken from the {\tt
TilingBoundary} table (which, itself, is a view of the {\tt
TilingGeometry} table with all the tiling masks removed) in the
DR5 CAS server. We reject those tiling rectangles with target
version lower than v3\_1\_0. The spectroscopic tile (plate)
information is taken from the {\tt maindr5spectro.par} table from
the DR5 website\footnote{{\tt http://www.sdss.org/dr5/}}, which
only includes tiles in the main survey and contains information of
which tiling chunk each tile belongs to. We create the sectors by
combining the two geometries using the spherical polygon
description in Hamilton \& Tegmark (2004). When computing the
effective area of either all the non-overlap sectors or all the
overlap sectors we use the {\it balkanization}
procedure in A. Hamilton's product {\em mangle}\footnote{{\tt http://casa.colorado.edu/$\sim $ajsh/mangle}} to reduce duplicate area.

After rejecting those tiling rectangles which used this earlier
version, our sample covers a solid angle of 4041 deg$^2$, of which
roughly 30\% is in overlap sectors. Because quasars in the overlap
regions are not subject to the restriction of not targeting pairs
separated by less than $55''$, and because the tiling algorithm
deliberately places the tile overlap in regions of higher target
density, one concern is that the angular selection function needs
to take into account a higher selection function in the overlap
region. However, we found that the number density of quasar
candidates (here looking at all redshifts, not just the
high-redshift candidates), and the number density of
spectroscopically confirmed quasars, were essentially identical in
the overlap and non-overlap sectors. In contrast, the number
density of spectroscopic galaxies in the overlap sectors (93.1
deg$^{-2}$) is 23\% higher than that in the non-overlap sectors
(75.4 deg$^{-2}$), due to the deliberate placing of the overlaps
in regions of high target density; galaxies dominate the SDSS
spectroscopic targets, and beyond a subtle effect due to
gravitational lensing (Scranton \etal\ 2005), we expect no
correlation between the background quasars and the foreground
galaxies.  All this means that the angular selection function of
our sample can be assumed to be uniform within the mask defined by
the sectors that make up our sample.
For DR5, the overall spectroscopic completeness of quasar
candidates is $\sim 95\%$, and the fraction of quasar candidates
that are indeed quasars is $\sim 48\%$.  The angular quasar number
density is $\sim 9.4$ deg$^{-2}$.

\section{C. Relationship between halo mass, clustering strength, and
  quasar lifetime}
\label{app:martini_weinberg}

In this appendix we follow Martini \& Weinberg (2001), and provide
some essential formulae to compute the quasar lifetime $t_{\rm Q}$ and
duty cycle using
the measured correlation length and quasar number density.

The Martin-Weinberg model is very sensitive to the halo number
density at the high mass end, hence a more suitable fitting
function is needed. The Press \& Schechter (1974; PS) halo number
density as a function of halo mass $M$ and redshift is given by:
\begin{equation}\label{n_M}
n(M,z)dM=-\sqrt{\frac{2}{\pi}}\frac{\rho_0}{M}\frac{\delta_c(z)}{\sigma^2(M)}\frac{d\sigma(M)}
{dM}\exp\left[-\frac{\delta_c^2(z)}{2\sigma^2(M)}\right]dM\ ,
\end{equation}
where $\rho_0=2.78\times 10^{11}\Omega_M\ h^2\ M_\odot\ {\rm
Mpc}^{-3}$ is the mean density of the universe at $z=0$;
$\sigma(M)$ is the current ($z=0$) rms linear density fluctuation
smoothed by a spherical top-hat with radius
$r=(\frac{3M}{4\pi\rho_0})^{1/3}$, normalized by $\sigma_8$; and
$\delta_c(z)=\delta_{c,0}/D(z)$ is the threshold density for
collapse of a homogeneous spherical perturbation at redshift z,
with $D(z)$ the growth factor and $\delta_{c,0}$ the critical
threshold at $z=0$, given in Appendix A of Navarro, Frenk, \&
White (1997). The Sheth-Tormen (ST) halo mass function is (Sheth
\& Tormen 1999)
\begin{equation}\label{n_M_ST}
n(M,z)dM=-A\sqrt{\frac{2a}{\pi}}\frac{\rho_0}{M}\frac{\delta_c(z)}{\sigma^2(M)}\frac{d\sigma(M)}
{dM}\left\{1+\left[\frac{\sigma^2(M)}{a\delta_c^2(z)}\right]^p\right\}\exp\left[-\frac{a\delta_c^2(z)}{2\sigma^2(M)}\right]dM\
,
\end{equation}
where $A=0.3222$, $a=0.707$ and $p=0.3$. We compared the ST and PS
formalism using the $z=3$ and $z=4$ outputs of a cosmological $N$-body
simulation generated from the TPM code of Paul Bode and Jeremiah P. Ostriker
(Bode, Ostriker, \& Xu 2000; Bode \& Ostriker 2003) which assumed the WMAP3
cosmology ($\Omega_m=0.26$, $\Omega_\Lambda=0.74$, $H_0=72\,\rm
km\,s^{-1}\,Mpc^{-1}$, spectral index $n_s=0.95$, and $\sigma_8=0.77$.
The simulation included $\sim 10^9$ particles in a box 1000 comoving
$h^{-1}$ Mpc on a side; the mass per particle was $6.72\times
10^{10}h^{-1}M_\odot$.  Dark matter halos were identified with the
Friends-of-Friends algorithm using a linking parameter one fifth of
the mean interparticle separation of the simulation.
We found that the mass function in the simulations for $M > 2
\times 10^{12} h^{-1}M_\odot$ followed the ST predictions closely,
while the PS form increasingly underpredicted the simulations at
large masses, in agreement with a number of other authors (e.g.,
Sheth \& Tormen 1999; Jenkins \etal\ 2001; Heitmann \etal\ 2006).
Therefore we use the ST formula for the halo mass function
throughout the paper.

The rms density fluctuation at $z=0$, $\sigma(M)$, is given by
\begin{equation}\label{sigma_M}
\sigma(M)=\left[\frac{1}{2\pi^2}\int_0^\infty
dk\,k^2P(k)\tilde{W}^2(kr)\right]^{1/2}\ ,
\end{equation}
where $\tilde{W}=3(\sin kr-kr\cos kr)/(kr)^3$ is the filter
function for a spherical top-hat. The CDM power spectrum
$P(k)\propto k^{n_s}T^2(k)$ where $n_s$ is the primeval
inflationary power spectrum index and $T(k)$ is the transfer
function, given by (Bardeen \etal\ 1986):
\begin{equation}
T(k)=\frac{\ln
(1+2.34q)}{2.34q}[1+3.89q+(16.1q)^2+(5.46q)^3+(6.71q)^4]^{-1/4}\ ,
\end{equation}
where $q=k/\Gamma$ and $\Gamma$ is the CDM shape parameter (with
units of $h\ {\rm Mpc}^{-1}$), given approximately by
$\Gamma=\Omega_Mh\exp [-\Omega_b-(2h)^{1/2}\Omega_b/\Omega_M]$
(Sugiyama 1995). Using this CDM power spectrum we numerically
integrate equation (\ref{sigma_M}) to obtain $\sigma(M)$ and
$d\sigma(M)/dM$. The rms fluctuation at redshift $z$ is thus given
by
\begin{equation}
\sigma(M,z)=\sigma(M)D(z)\ ,
\end{equation}
from which we can define a characteristic mass scale $M_*$, such that
$\sigma[M_*(z)]=\delta_c(z)$.

The halo lifetime is defined to be the median interval before a
halo with initial mass $M$ becomes a halo with mass $M_2=2M$ via
mergers. This condition is given in Lacey \& Cole (1993),
\begin{equation}\label{t_halo}\footnotesize
P(S<S_2,\omega_2\mid S_1,
\omega_1)=\frac{1}{2}\frac{\omega_1-2\omega_2}{\omega_1}\exp\left[\frac{2\omega_2(\omega_1-\omega_2)}{S_1}
\right]{\rm
erfc}\left[\frac{S_2(\omega_1-2\omega_2)+S_1\omega_2}{\sqrt{2S_1S_2(S_1S_2)}}\right]
+\frac{1}{2}{\rm
erfc}\left[\frac{S_1\omega_2-S_2\omega_1}{\sqrt{2S_1S_2(S_1-S_2)}}\right]=0.5\
,
\end{equation}
where $S_1=\sigma^2(M)$, $S_2=\sigma^2(2M)$,
$\omega_1=\delta_c(z)$ and $\omega_2=\delta_c(z_2)$. Hence the
halo lifetime is given by $t_{\rm H}(M,z)=t_{\rm U}(z_2)-t_{\rm
U}(z)$, where $t_{\rm U}(z)$ the age of the universe at redshift
$z$, and $z_2$ is solved numerically from eqn. (\ref{t_halo}).
For comparison, the age of the universe at $z=3.5$ is $\sim 2$ Gyr.

Halos with mass $>M_*$ are more strongly clustered than the
underlying distribution of mass; the bias factor $b(M,z)$ of halos
with mass $M$ at redshift $z$ is given by (Jing 1998)
\begin{equation}\label{bias}
b(M,z)=\left\{1+\frac{1}{\delta_{c,0}}\left[\frac{\delta_c^2(z)}{\sigma^2(M)}-1\right]\right\}
\left[\frac{\sigma^4(M)}{2\delta_c^4(z)}+1\right]^{(0.06-0.02n_{\rm
eff})}\ ,
\end{equation}
where $n_{\rm eff}=-3-6(d\ln \sigma/d\ln M)$ is the effective
index of the power spectrum on a mass scale $M$.
The effective bias factor for all halos with mass above the
minimal mass $M_{\rm min}$ is therefore
\begin{equation}\label{b_eff}
b_{\rm eff}(M_{\rm min},z)=\int_{M_{\rm min}}^\infty
dM\frac{b(M,z)n(M,z)}{t_{\rm H}(M,z)}\left[\int_{M_{\rm
min}}^\infty dM\frac{n(M,z)}{t_{\rm H}(M,z)}\right]^{-1}\ .
\end{equation}
Since $n(M,z)$ drops rapidly with increasing mass, $b_{\rm eff}$
is only slightly larger than $b(M_{\rm min},z)$. We have tested
equations~(\ref{bias}) and (\ref{b_eff}) with the simulations
described above. We find that they correctly predict the bias
inferred from the integrated correlation function $\xi_{20}$. In
particular, at the two output redshifts of the simulations, $z=3$
and $z=4$, the simulation results give a bias factor (calculated
from the ratio of $\xi_{20}$ for the halos and for the dark
matter) of $6.2$ at $z=3$ and 10.2 at $z=4$, for halos with mass
$\ge 2\times 10^{12}\ h^{-1}\ M_\odot$; while the analytical bias
formalism gives $b_{\rm eff}=7.3$ and $10.7$ respectively. This
difference is negligible when we integrate over a wide redshift
range (equation~\ref{xi_average}) and compared with other
uncertainties. On the other hand, there is clear evidence for a
scale-dependent bias, which we plan to explore further in future
work.

The model predicted quasar correlation function $\xi_{\rm
model}(r,z)$ is therefore
\begin{equation}\label{xi_model}
\xi_{\rm model}(r,z)=b_{\rm eff}^2\xi_m(r,z)=b_{\rm
eff}^2\xi_m(r)D^2(z)\ ,
\end{equation}
where $D(z)$ is the linear growth factor of fluctuations, and
$\xi_m(r)$ is the present-day mass correlation function, defined
as
\begin{equation}
\xi_m(r)=\frac{1}{2\pi^2}\int_0^\infty dk\,k^2P(k)\frac{\sin
kr}{kr}\ ,
\end{equation}
normalized using $\sigma_8$.  Comparison of $\xi_m(r,z)$ with
the mass correlation function from the cosmological N-body
simulation mentioned above at $z=3$ and $z=4$ shows quite good
agreement.

The correlation function we have actually measured is averaged over a
certain redshift range, hence
\begin{equation}\label{xi_average}
\bar{\xi}(r)=\frac{\int dV_c\ n_{\rm QSO}^2(z)\xi_{\rm
model}(r,z)}{\int dV_c\ n_{\rm QSO}^2(z)}\ ,
\end{equation}
where $n_{\rm QSO}(z)=\Phi(z)f(z)$ is the {\em observed} quasar number
density, i.e., the actual quasar number density times the
complicated selection function $f(z)$; and $dV_c$ is the
differential comoving volume element, given in Hogg (1999).
$n_{\rm QSO}$ is computed using our full high-redshift clustering
subsample; see Figure~\ref{fig:z_distri}. Note that the above equation is only valid for scales
$r$ over which $n_{\rm QSO}$ is near constant and $\xi$ does not
significantly evolve over the time $r/[(1+z)c]$ (PMN04). For our
selected range $[r_{\rm min}, r_{\rm max}]=[5,20]\, h^{-1}$ Mpc,
these conditions are satisfied.

\end{document}